\documentclass[lettersize,journal, 10pt]{IEEEtran}
\usepackage{amsmath,amsfonts}
\usepackage{algorithmic}
\usepackage{academicons}
\usepackage{xcolor}
\definecolor{orcidlogocol}{HTML}{A6CE39}
\usepackage{algorithm}
\usepackage{array}
\usepackage[caption=false,font=footnotesize,labelfont=sf,textfont=sf]{subfig}
\usepackage{textcomp}
\usepackage{stfloats}
\usepackage{url}
\usepackage{verbatim}
\usepackage{graphicx}
\usepackage[hidelinks]{hyperref}
\usepackage{cite}
\usepackage{multirow}
\usepackage{booktabs}
\usepackage{enumitem}
\usepackage{scalerel}
\usepackage{tikz}
\usetikzlibrary{svg.path}

\definecolor{orcidlogocol}{HTML}{A6CE39}
\tikzset{
  orcidlogo/.pic={
    \fill[orcidlogocol] svg{M256,128c0,70.7-57.3,128-128,128C57.3,256,0,198.7,0,128C0,57.3,57.3,0,128,0C198.7,0,256,57.3,256,128z};
    \fill[white] svg{M86.3,186.2H70.9V79.1h15.4v48.4V186.2z}
                 svg{M108.9,79.1h41.6c39.6,0,57,28.3,57,53.6c0,27.5-21.5,53.6-56.8,53.6h-41.8V79.1z M124.3,172.4h24.5c34.9,0,42.9-26.5,42.9-39.7c0-21.5-13.7-39.7-43.7-39.7h-23.7V172.4z}
                 svg{M88.7,56.8c0,5.5-4.5,10.1-10.1,10.1c-5.6,0-10.1-4.6-10.1-10.1c0-5.6,4.5-10.1,10.1-10.1C84.2,46.7,88.7,51.3,88.7,56.8z};
  }
}

\newcommand\orcidicon[1]{\href{https://orcid.org/#1}{\mbox{\scalerel*{
\begin{tikzpicture}[yscale=-1,transform shape]
\pic{orcidlogo};
\end{tikzpicture}
}{|}}}}

\usepackage{hyperref} 

\begin{document}

\title{DCO: Dynamic Cache Orchestration for LLM Accelerators through Predictive Management}

\author{Zhongchun~Zhou\orcidicon{0009-0000-7037-7418},~Chengtao~Lai\orcidicon{0000-0002-9547-9653},~Yuhang~Gu\orcidicon{0009-0001-5040-9424},~and~Wei~Zhang\orcidicon{0000-0002-7622-6714},~\IEEEmembership{Fellow,~IEEE}%
\thanks{This work was supported by ZTE Industry-University-Institute Cooperation Funds under Grant No. IA20250923007, and was also partially funded by the Hong Kong Research Grants Council under Grant GRF16219825.}%
\thanks{A preliminary version of this paper appeared in the Proceedings of the 38th ACM International Conference on Supercomputing (ICS), 2024~\cite{LCM}.}%
\thanks{Z. Zhou and C. Lai contributed equally to this work.}%
\thanks{Z. Zhou, C. Lai, and Y. Gu are with the Department of Electronic and Computer Engineering, 
The Hong Kong University of Science and Technology, Clear Water Bay, Kowloon, Hong Kong 
(e-mail: zzhouch@connect.ust.hk; claiaf@connect.ust.hk; ygubp@connect.ust.hk).}%
\thanks{W. Zhang (corresponding author) is with the Department of Electronic and Computer Engineering, 
The Hong Kong University of Science and Technology, Clear Water Bay, Kowloon, Hong Kong 
(e-mail: eeweiz@ust.hk).}%
}

\IEEEpubidadjcol

\maketitle
\begingroup
\renewcommand\thefootnote{}
\addtocounter{footnote}{-1}
\endgroup
\begin{abstract}
The rapid adoption of large language models (LLMs) is pushing AI accelerators toward increasingly powerful and specialized designs.
Instead of further complicating software development with deeply hierarchical scratchpad memories (SPMs) and their asynchronous management,
we investigate the opposite end of the design spectrum: a multi-core AI accelerator equipped with a shared system-level cache and application-aware management policies, while keeping the programming effort modest.
Our approach exploits dataflow information available in the software stack to guide cache replacement (including dead-block prediction), in concert with bypass decisions and mechanisms that alleviate cache thrashing.

We evaluate the proposal using a cycle-accurate simulator on FlashAttention-based LLM workloads and observe substantial performance gains—up to 1.90× speedup over a conventional LRU-managed LLC across a range of cache sizes and dataflows.
In addition, we build and validate an analytical model that takes into account the actual overlapping behaviors to extend the measurement results of our policies to real-world larger-scale workloads. Experimental results show that when functioning together, our dead block prediction, bypassing, and thrashing mitigation strategies can handle scenarios both with and without inter-core data sharing and achieve remarkable speedups.
Finally, we implement the TMU in RTL using Chisel and synthesize it in a 15 nm standard-cell library: the TMU occupies only $\mathbf{0.064mm^2}$ and can operate at 2 GHz. Our findings demonstrate that a shared system-level cache, when coupled with lightweight tensor-aware orchestration, is a practical and scalable alternative to deeply hierarchical software-managed SPMs in future AI accelerator systems.

\end{abstract}

\begin{IEEEkeywords}
AI accelerator, analytical model, large language model, memory management.
\end{IEEEkeywords}

\section{Introduction}
\label{intro}
\IEEEPARstart{S}{ince 2014}, extensive efforts in academia and industry have led to the development of deep learning accelerator prototypes and chips, such as Eyeriss\cite{chen16}, Gemmini\cite{gemmini}, and industry examples like Google's TPUs\cite{tpu}, and Huawei's Ascend 310 and 910\cite{davinci}, which meet the demand from data centers \cite{tccl,hpn} to embedded systems. 
\IEEEpubidadjcol

AI accelerators, compared with Graphics Processing Units (GPUs), can be optimized for AI applications and tailored for specific scenarios, such as pre-defined neural network (NN) computation graphs, operator types, certain data precision, and given power budgets.
Since they are often used in scenarios where the execution graph is known during compilation,
they typically employ software-controlled scratchpad memories (SPMs) as the on-chip storage.
SPMs, similar to caches, are also Static Random-Access Memories (SRAMs), but lack address tags and status bits. This design choice undoubtedly simplifies hardware design, leaves more chip area for computing units and on-chip storage, and enables higher performance due to asynchrony.  However, these benefits also come with costs.
In addition to SPM data fragmentation that needs to be resolved during graph compilation \cite{tela},
data transfer between SPMs and the main memory needs to be scheduled explicitly by programmers or compilers. Such transfer primitives are often designed to be asynchronous to enable computation-communication overlap, so barriers are also required.
Simple as they seem, programming will quickly become prohibitive when combined with some other features: (1) when multiple levels of on-chip buffers exist (e.g., Cambricon's BANG-C programming model requires a 5-level pipeline and $>200$ lines of code only to implement a high-performance vector add \cite{cambricon}), 
(2) when the programming model is SIMD-based (Single Instruction Multiple Data) which requires handling unaligned cases and tail blocks carefully.
Hence, \textbf{to alleviate these programming complexities}, our work explores the potential of a \textbf{hybrid architecture that replaces shared SPM hierarchy with a last-level cache} (LLC) in AI accelerators.

This choice is not uncommon, as is demonstrated by the extensive use of
NVIDIA 
GPUs in AI workloads. However, the impact of caches on large-scale AI accelerator systems remains underexplored. In this work, we examine the role and strategies of cache management in the context of deep learning and LLMs. Our contributions include:
\begin{itemize}
\item We propose a Tensor Management Unit (TMU) integrated into the memory hierarchy to assist shared-cache management in multi-core AI accelerators. TMU maintains lightweight tensor- and tile-level metadata and exposes predictive hints to the LLC controller. We implement TMU in Chisel HDL and synthesize it in a 15 nm process library to quantify its practical overhead.
\item Based on metadata stored in TMU, we design a self-adaptive anti-thrashing mechanism, a dead-block prediction scheme, and a dynamic bypassing policy that work cooperatively to help the cache capture reuses within a large working set.
\item We establish an analytical model through bottleneck and overlapping analysis, and validate it against our cycle-level simulator. This enables us to evaluate our strategies under real-world larger-scale cases.
\end{itemize}

A preliminary version of this work appeared at ICS 2024 as LCM: LLM-focused Hybrid SPM-cache Architecture with Cache Management for Multi-Core AI Accelerators~\cite{LCM}. We substantially extend that work by redesigning the TMU to cut hardware area by 65\%, and by introducing a unified priority scheme that coordinates a new dynamic cache bypassing mechanism with our anti-thrashing replacement policy. We further develop a new analytical model validated against cycle-accurate simulation, and reevaluate the architecture using GQA FlashAttention workloads (e.g., Llama3), demonstrating the effectiveness of our framework on modern LLM operators.

\section{Background and Motivation}
Below, we first briefly review the basics of Transformers and FlashAttention, which then lead to a short discussion on existing GPU/AI accelerator architectures and their limitations. After that we show the factors that motivate us to incorporate a system-level cache in place of a shared SPM in AI accelerators, as well as how our proposal aligns with current trends:
(1) (explained in Section \ref{intro}) shared SPM management becomes increasingly complex in conventional architecture, while using a shared cache can ease development; (2) System-on-Chips (SoCs) in cell phones and autonomous vehicles usually involve a system-level cache, so we also try gaining insights on using it for LLMs without introducing additional large buffers; (3) Our work is an example of exposing tensor-level information to the hardware, a trend already present in current GPUs.

\subsection{Transformer and FlashAttention}

Due to the quadratic growth of intermediate tensors ($\mathbf{QK}^T$) with respect to sequence length, self-attention would normally exhibit a low compute-to-memory-access ratio, resulting in significant under-utilization of computing units.
As a widely accepted solution, FlashAttention \cite{dao2022} leverages online softmax to store and update each attention score tile on-chip, hence eliminating the need to materialize it in the global memory.

As the GPU architecture evolves, FlashAttention has also experienced several major updates. FA-2 \cite{flashattention2} better utilizes the Tensor Cores, modifies the dataflow to better parallelize the thread blocks among stream multiprocessors (SM), and improves work partition between different warps of a thread block to reduce shared memory access. FA-3 \cite{fa3} further exploits features in Hopper GPUs: asynchronous execution of Tensor Cores, CUDA Cores and Tensor Memory Accelerator (TMA), as well as enhanced low-precision capabilities.

However, little effort has been made from a general architecture's perspective that does not require completely rewriting operator programs. TMA and warp specialization are unique to recent NVIDIA GPUs and may not see their counterparts in other GPUs or NPUs. Due to the deep software pipelines, FA-3 can be register-bound on Hopper GPUs, but this may not be the case for other devices. In other words, the performance bottleneck shifts with system resources and operator implementation. To broaden the scope of this work, we only augment the shared cache with some universal strategies and do not make other specific architectural assumptions.

\subsection{AI Accelerators in SoCs and Existing Architectural Support}
\label{SoC}
We focus on multi-core accelerators with a shared last-level cache. Such designs are preferred over single-core embedded ones like Eyeriss-style chips~\cite{chen16}, because LLMs are typically deployed on multi-core or multi-accelerator systems. Previous works have also shown downgraded scalability in terms of performance and energy efficiency through increasing computing and memory resources inside a single AI core~\cite{TANGRAM, simba}.

An increasing number of commercial SoCs already integrate AI 
accelerators, and expose a shared system-level 
cache to all compute agents. Examples include mobile SoCs such as the 
Qualcomm Snapdragon X Elite series~\cite{xelite} and ARM’s DynamicIQ-based 
designs~\cite{SoCdesign}, where a common LLC is used to reduce dark silicon and 
improve overall chip efficiency. In such systems, the system-level 
cache is effectively irreplaceable by a private scratchpad, because it 
also serves CPUs and other agents. This makes “how to use an existing 
shared cache for AI/LLM workloads” a more realistic question than 
“whether to add a new large global buffer”.



In addition to a shared cache, recent industrial designs have begun to move structured tensor metadata closer to the hardware. For instance, NVIDIA’s Hopper architecture introduces TMA engines and tensor maps that encode multi-dimensional addresses, strides, and layouts; these descriptors are consumed by specialized load instructions to orchestrate asynchronous tile transfers into shared memory. Similarly, Triton-based kernels organize tensor accesses around lightweight descriptor objects (i.e., \texttt{triton.language.} \texttt{make\_tensor\_descriptor}), which capture base pointers, shapes, and strides to generate TMA instructions.

These mechanisms show that exposing tensor-level information (shape, layout, tiling) to the hardware is already a widely adopted 
practice in high-performance kernels. However, such metadata 
is typically confined to driving local SRAM transfers and kernel-specific software pipelines, failing to systematically
coordinate the behavior of a shared LLC across multiple cores.

Our work can be viewed as generalizing this trend from 
per-kernel tensor descriptors to a reusable architectural 
primitive for shared-cache management. Instead of embedding 
tensor metadata only inside individual kernels, we introduce 
a small Tensor Management Unit that maintains per-tensor 
and per-tile information and feeds it to the LLC replacement 
and bypass logic. This allows the shared cache to make 
predictive decisions—such as dead-block identification, 
anti-thrashing prioritization, and dynamic bypassing—based 
on the same type of high-level tensor descriptors that are 
already common in modern LLM software stacks.

In summary, existing SoCs increasingly provide a shared 
system-level cache that is already on the critical path of AI/LLM workloads, and industrial software/hardware stacks 
have begun to expose tensor descriptors to drive 
fine-grained data movement. Building on these trends, our 
work studies how to orchestrate a shared LLC among multiple 
AI cores using lightweight tensor-level metadata, rather than relying solely on software-managed SPMs. Our DCO mechanisms do 
not assume any specific AI core microarchitecture; instead, they only require that cores issue bulk transfers between their private SPMs and the shared cache, making the proposal applicable to a wide range of GPU- and NPU-like systems.

\label{SoCDesign}
\section{System Architecture Overview}
\subsection{Baseline SoC Integration Options} 
Integrating AI accelerators into the SoC becomes a common practice, but the architectural designs and data sharing mechanism between CPU and AI cores vary. Previous works typically assume an SoC like Fig. \ref{fig:AssumptionStructure}(b)-(d). (b) and (c) are solely SPM-based. (d) guides accelerator traffic to the CPU L2 cache, relieving the compiler from shared SPM allocation, while still capturing data reuse between AI cores. Despite the existence of alternative methods such as \cite{fang2015exploring} to mitigate the impact of substantial accelerator traffic on CPU execution, the inherent CPU-NPU intervention remains a persistent challenge.

\subsection{Proposed DCO-Based SoC Architecture}
To mitigate cache pollution imposed by CPUs, we adopt the configuration depicted in Fig. \ref{fig:AssumptionStructure}(e). 
This architecture retains the advantage of Fig. \ref{fig:AssumptionStructure}(d): no need for shared buffer allocation and thus more friendly to the compiler.
Note (e) is an architectural abstraction where CPU and accelerator traffic are separated. For a real SoC, the logical behavior can approximate either (d) or (e) depending on whether CPU traffic is configured to bypass the accelerator-shared cache.
This shared cache architecture among AI cores resembles that of a GPU, where each core with its private SPM corresponds to an SM.

In order to isolate variables and avoid the non-deterministic requests issued by the CPU cores, we consider a dedicated accelerator LLC in our experiments. However, DCO is compatible with a shared LLC used in modern SoCs (e.g., Apple Silicon). Memory requests from CPUs will simply miss in TMU and fall back to the default cache replacement policy like LRU.
From the accelerator's viewpoint, concurrent CPU traffic then manifests itself
as a reduction of the LLC capacity effectively available to AI tensors. Our design performs well when the capacity of LLC is limited according to Section~\ref{sec:exp}.

\subsection{Challenges for Shared Caches in Multi-core AI Systems}
The performance of a cache-based system is sensitive to the hit rate, which is closely related to the working set.
Furthermore, Transformer and FlashAttention
kernels would put the shared cache under stress, because the working set for long-context LLMs could far exceed the LLC capacity. A naive LRU policy will degenerate into thrashing: useful lines are frequently evicted before their next use, so the hit rate becomes remarkably low or even 0.
The observation thus motivates us to design a replacement and bypass policy that can capture as much reuse as possible from an oversized working set.

\subsection{Role of the Tensor Management Unit (TMU)}

\begin{figure}[t]
\vspace*{-0.3\baselineskip}
\centerline{\includegraphics[width=1\linewidth]{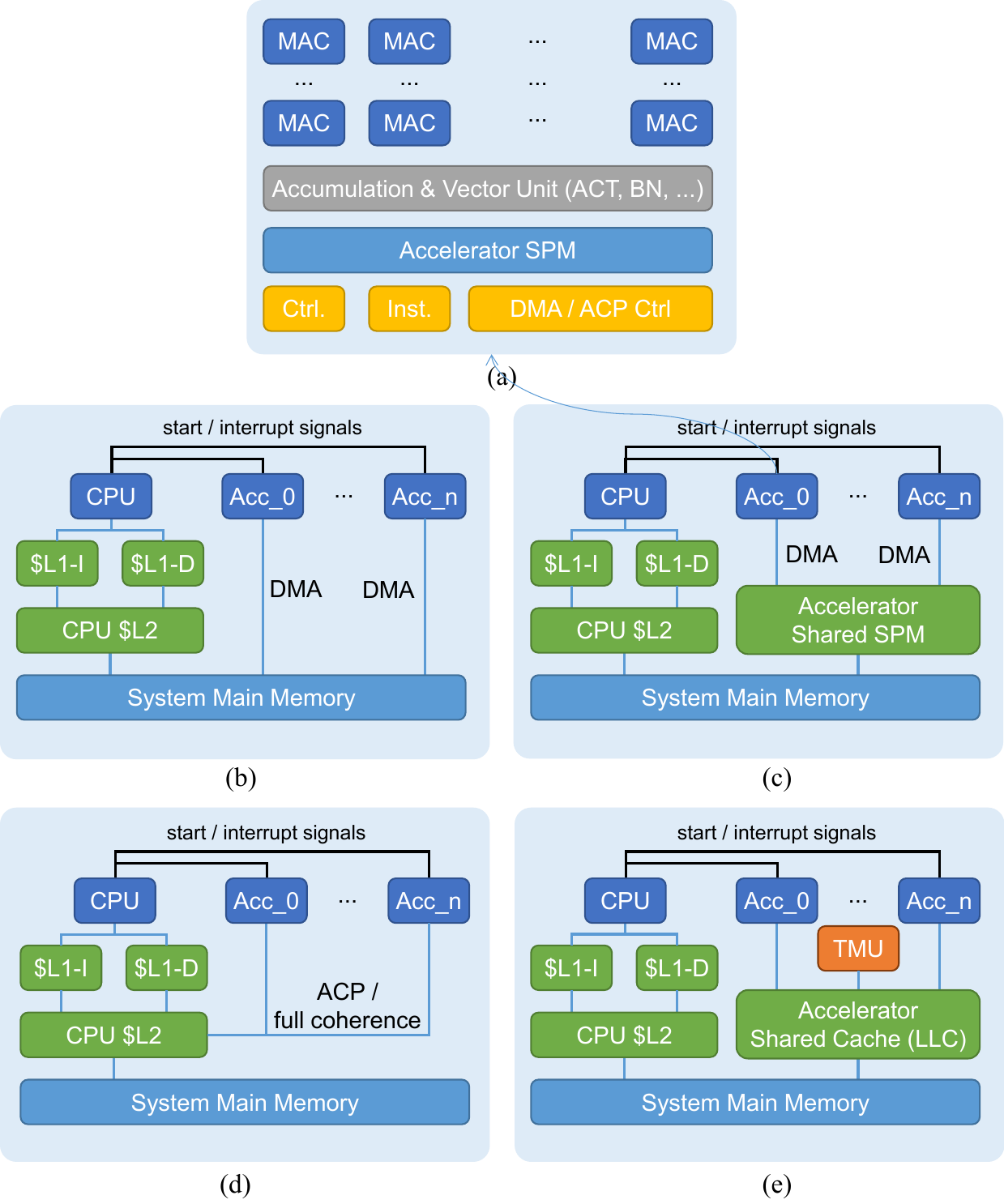}}
\vspace*{-.7\baselineskip}
\caption{(a) A typical AI accelerator structure. (b) accelerators directly connected to DRAM via DMA. (c) accelerators connected to a shared SPM. (d) accelerators connected to the CPU L2 cache. (e) (\textbf{ours}) accelerators first connected to a shared cache (aided by an additional ``TMU'').\cite{LCM}}
\label{fig:AssumptionStructure}
\vspace{-5mm}
\end{figure}
To address the above challenges, a replacement and bypass policy alone is not sufficient; it also needs high-level information about tensor lifetimes and dataflow that is typically only known to software. Meanwhile, only the hardware has real-time visibility into resource usage and cache pressure.
With this in mind, we propose to provide the hardware with workload information that is traditionally only available on the software side. We aid the shared LLC with auxiliary data structures that hold information about the workload. We have also added additional hardware units to react in different situations with the help of such data structures. The additional unit (TMU) is shown in orange in Fig. \ref{fig:AssumptionStructure}(e).
The cache’s replacement and bypass logic then consults
lightweight TMU-provided hints to implement dead block prediction (predicting cache lines that have finished their lifespans), anti-thrashing,
and dynamic bypassing. The detailed microarchitecture and
metadata formats are described in Section~\ref{dbp_sec}.

\section{Mixed-Strategy Cache Management for Deep Learning Workloads}
\label{dbp_sec}
\subsection{Dead Block Prediction and Anti-thrashing}
\label{MainIntro}
As two perspectives to enhance performance, dead block prediction and anti-thrashing strategies have been widely discussed for conventional CPU architectures and workloads, but still require specialized solutions for AI accelerators running data-intensive workloads like Transformers.

Replacement policies like LRU generate victims according to the access history. But the ranks of cache lines given by such policies do not necessarily reflect their life cycles.
Fortunately, for a given dataflow, the number of reuses can be precisely known before computation, making a counter-based method feasible.
Fig. \ref{fig:TMUStruct}(a) provides a conceptual example of registering necessary information into the hardware via specialized instructions.
The compiler pass runs after tiling and before instruction generation (e.g., on
Triton-IR or MLIR \texttt{linalg}), where tile-level index maps are known. For
a tensor $T$ accessed in a loop nest $L$ with trip counts $N_l$, let
$V_T\!\subseteq\!L$ be the loops whose induction variables appear in the tile
index map of $T$; the tile is invariant along every $l\!\notin\!V_T$, hence
$\mathrm{nAcc}(T)=\prod_{l\in L\setminus V_T}N_l$. In Fig.~\ref{fig:TMUStruct}(a)
this gives $J$, $I$ and $1$ for $A$, $B$ and $C$---exactly the registered
values. \texttt{tile\_size} and \texttt{id} follow from the tile extents and
the operand position, and \texttt{bypass} is asserted when
$\mathrm{nAcc}\!=\!1$, which is how $Q$/$O$ of FlashAttention are bypassed
automatically. \texttt{reg}/\texttt{clear} are hoisted to the kernel
prologue/epilogue, where base addresses and symbolic trip counts are resolved,
so dynamic shapes need no recompilation.


We integrate our solution into the replacement policy. When searching for a victim, it first tries to find a dead block. If no dead block is found, it turns to the anti-thrashing policy, which keeps a certain subset of the working set in the cache. If this procedure produces a tie, it falls back to LRU to select the final victim within those anti-thrashing candidates. Straightforward as it is, the hardware should be implemented with special optimizations to reduce area and latency overhead.

\subsection{Hardware for Dead Block Prediction and Anti-thrashing}
\label{DBPDetail}
Section \ref{MainIntro} demonstrates the necessity of using dataflow-related information of cache lines for a better replacement policy. In our design, this is achieved by introducing an additional hardware unit named tensor management unit (TMU). TMU acts as a liaison between software and hardware: metadata of tensors and tiles are written into TMU with specialized instructions after allocation and before operator computation, so dynamic shapes are also supported; the metadata are later retrieved by the replacement policy to get the victims. These metadata are managed at a \textbf{tensor- and tile-level instead of a cache-line level} since cache lines in the same tile share the same properties. A coarser granularity also reduces resource usage since it requires fewer entries. TMU parameters are listed in Table \ref{table:vars}, and will be explained upon reference.

\begin{scriptsize}
\begin{table*}[t]
  \centering
  \vspace*{-0.6\baselineskip}
  \caption{Variables used in the proposed architecture}
  \vspace*{-0.4\baselineskip}
  \label{table:vars}
  \begin{tabular}{l|l|l}
    \hline
    \multicolumn{1}{c|}{Symbol} & \multicolumn{1}{c|}{Meaning} & \multicolumn{1}{c}{Related Criteria}\\
    \hline
    \texttt{D\_LSB} & least significant bit to select bits from tag when looking up \texttt{dead\_fifo} & \multirow{2}{*}{\begin{tabular}{@{}l@{}} considered as dead if \texttt{tag[D\_MSB:D\_LSB] \textbf{is in}} \\
    \texttt{dead\_fifo} \end{tabular}}\\
    \cline{1-2}
    \texttt{D\_MSB} & most significant bit to select bits from tag when looking up \texttt{dead\_fifo} & ~\\
    \hline
    \texttt{B\_BITS} & bits to select from tag when deciding bypass; also used in anti-thrashing & \multirow{2}{*}{\begin{tabular}{@{}l@{}} bypass if \texttt{tag[B\_BITS-1:0] < B\_GEAR}; \\
    anti-thrash: first evict lines with smaller \texttt{tag[B\_BITS-1:0]}\end{tabular}} \\
      \cline{1-2}
  \texttt{B\_GEAR} & threshold to determine whether to bypass a cache line & ~ \\
  \hline
  \texttt{accCnt} & number of accesses of a certain cache line, updated upon each access & \multirow{2}{*}{\begin{tabular}{@{}l@{}} move a tile identifier from \texttt{live tile info} to \\ \texttt{dead tile identifiers} if \texttt{accCnt == nAcc} \end{tabular}} \\
  \cline{1-2}
  \texttt{nAcc} & expected number of accesses of each cache line of a tensor & ~ \\
  \hline
  \end{tabular}
  \vspace*{-0.4\baselineskip}
\end{table*}
\end{scriptsize}

\begin{figure*}[!t]
\centerline{\includegraphics[width=1\linewidth]{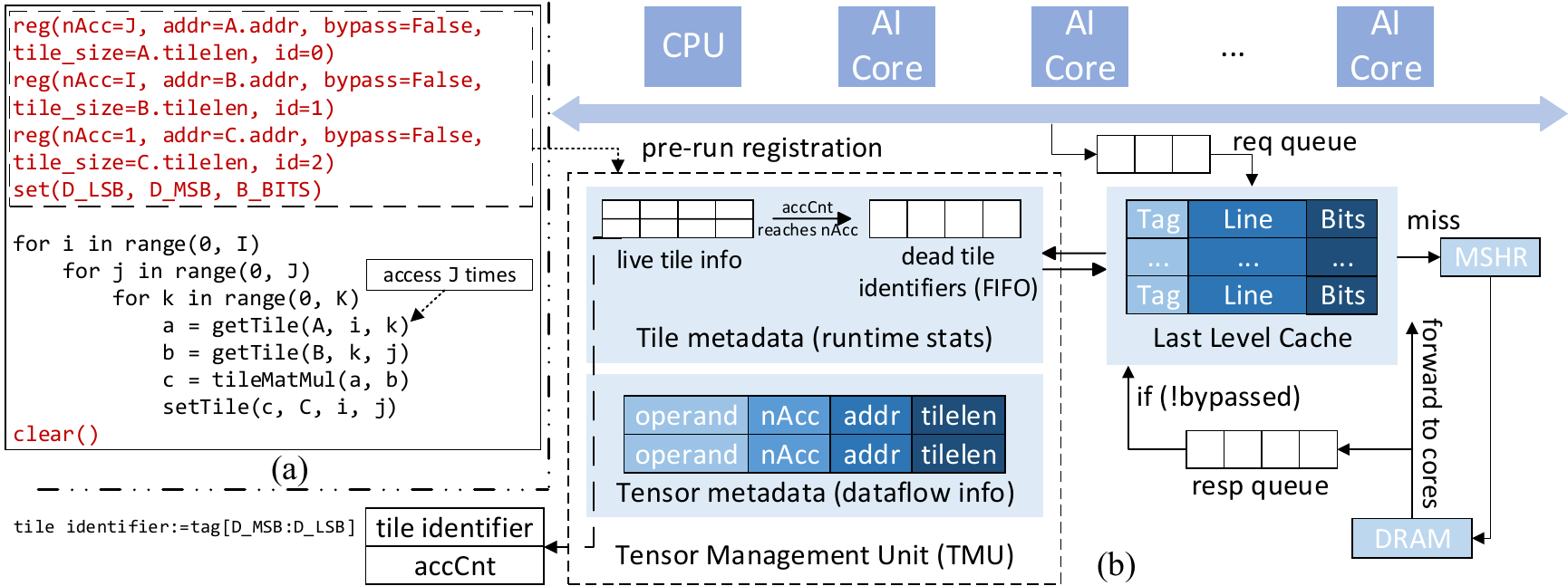}}
\vspace*{-0.2\baselineskip}
\caption{(a) Pseudo code of tiled MatMul. nAcc: expected number of accesses of each tile in this matrix; getTile: load the tile into the AI core private SPM; setTile: store the tile into the memory system from the AI core private SPM. Lines in red are manually-inserted registration codes to pass dataflow-related information to TMU. (b) System architecture and workflow with TMU. The interaction between LLC and TMU is elaborated in Section \ref{DBPDetail}.\cite{LCM}}
\label{fig:TMUStruct}
\vspace{-1\baselineskip}
\end{figure*}

Fig. \ref{fig:TMUStruct}(b) shows the internal structure of TMU and how it interacts with the rest of the system. TMU is made up of two components: the tensor metadata module and the tile metadata module. The former keeps static operator metadata that are registered before execution, while the latter consists of live tile info and dead tile identifiers, both of which assist cache replacement and are updated during runtime using tensor metadata. \texttt{accCnt} in the live tile info module will be updated when the \underline{t}ile's \underline{l}ast cache \underline{l}ine (TLL) is accessed.
Three specialized instructions are designed for the CPU to register dataflow-related information into the TMU:
\begin{itemize}
    \item Tensor metadata: Expected \texttt{numAccess} (determined by dataflow, a.k.a. \texttt{nAcc}), tensor base address, whether to bypass the whole tensor from LLC, tile size of the tensor, and operand id (e.g, left operand / right / output);
    \item Clear: Clear the current registration if no longer needed;
    \item Set other parameters: \texttt{D\_LSB}, \texttt{D\_MSB}, \texttt{B\_BITS}.
\end{itemize}

During accelerator computation, the TMU works as follows:
\begin{itemize}
    \item When an AI core requests an \texttt{addr}, the LLC informs TMU of this access. If it is a TLL, \texttt{accCnt} increases by 1. If it reaches the expected \texttt{nAcc}, the entry will retire from the live tile info module, and the bits \texttt{tag[D\_MSB:D\_LSB]} will be recorded into the dead tile identifier FIFO;
    \item  The size of the dead tile identifier FIFO is designed to be limited to achieve a short access latency so that replacement decisions can be made within a clock cycle. When the FIFO is full, an old element will be cleared.
    \item When the replacement policy needs to select a victim, it queries the FIFO whether the set contains a dead block. This information along with the anti-thrashing policy (Section~\ref{sec:sa_thrash}) will be used for the replacement policy.
\end{itemize}

Both directions of this interface are narrow and off the critical path.
(i)~\emph{Updates} (LLC$\rightarrow$TMU) carry only a slice id and the
\texttt{tag[D\_MSB:D\_LSB]} field---tens of bits, against a 64\,B line---and
are issued in parallel with the tag/data array access. Because they are
\emph{hints}, a delayed or dropped update only shifts \emph{when} a tile is
declared dead and never affects correctness, so neither stalling nor
serialization with the LLC pipeline is required. Furthermore, \texttt{accCnt}
is touched only on a TLL access: with 16\,kB tiles and 64\,B lines only 1 out
of 256 accesses is a TLL, so all 32 slices together generate at most
$32/256\!=\!0.125$ updates per cycle and a single-ported TMU is never a
throughput bottleneck. (ii)~\emph{Queries} (TMU$\rightarrow$LLC) are issued
only when a victim has to be chosen, i.e.\ on a fill rather than on every
access; the 16-entry dead-tile FIFO is a small CAM whose match resolves within
one cycle at 2.0\,GHz (Section~\ref{sec:cost}), so an eviction decision still
completes in a single cycle and the baseline LLC pipeline is left unmodified.



\subsection{Self-adaptive Anti-thrashing}
\label{sec:sa_thrash}


The anti-thrashing policy prioritizes part of the cache lines while sacrificing the others.
Rather than manually tune the management parameters according to cache size by engineers, we design a scheme with a self-adaptive nature.

We use a segment in the cache tag, \texttt{tag[B\_BITS-1:0]}, to decide the relative priorities of cache lines. These are the lowermost bits of the tag domain, so we expect this value to span a uniform distribution across a tensor. We let the replacement policy evict cache lines with smaller \texttt{tag[B\_BITS-1:0]} values first. If the result produces a tie, it will use the default LRU result. The smaller the cache is, the fewer cache lines it will keep, and the more it will sacrifice to mitigate thrashing.


To accommodate varying requirements in diverse applications, we can use different \texttt{B\_BITS} values. 
A larger \texttt{B\_BITS} means a finer control over cache thrashing but also leaves smaller space for LRU, and vice versa. The detailed experiments and analysis are in Section~\ref{design_choices}.



\subsection{Coordinated Dynamic Bypassing}

\label{sec:coordinated_bypass}

Even a well-designed replacement policy, as described in Section~\ref{DBPDetail} and~\ref{sec:sa_thrash}, can be overwhelmed when the working set is significantly larger than the cache. To further mitigate this, we augment the anti-thrashing mechanism with a dynamic bypassing policy that works collaboratively. The connection between them is a unified priority scheme, i.e., using the same least significant bits of an address tag as a priority score (\texttt{tag[B\_BITS-1:0]}) to control their decisions.
This synergy is crucial to avoiding conflicts between them.

\begin{itemize}

\item \textbf{Replacement Policy:} During an eviction, the anti-thrashing mechanism selects a victim from the lowest-priority tier currently present in the cache set. This approach protects higher-priority data, helping to maintain a smaller working set under pressure.

\item \textbf{Bypass Policy:} Upon a cache miss, if the incoming line has a low priority score, it may be a candidate to bypass LLC. This preemptive action spares the cache from allocating space for data that would likely be a primary eviction candidate anyway.

\end{itemize}


Like anti-thrashing, our bypassing policy can also adapt itself to the varying memory pressure at runtime.
This is achieved via a runtime-adaptive threshold \texttt{B\_GEAR}, which is managed independently in each cache slice.  This gear decides how many of the lowest-priority tiers are subject to bypassing.  The system dynamically tunes this value based on observed cache contention, measured via the number of cache evictions in an LLC slice during a recent time frame (LLC eviction rate). This feedback loop operates as follows:

\begin{itemize}

\item If the eviction rate surpasses a threshold (\texttt{bypass\_ub}), it signals severe cache thrashing.  The system responds by increasing \texttt{B\_GEAR}, thereby bypassing more low-priority data to alleviate pressure.

\item Conversely, if the eviction rate falls below a threshold (\texttt{bypass\_lb}), it indicates that the cache has ample capacity. The system then decreases \texttt{B\_GEAR}, allowing more data to be cached to maximize the potential hit rate.

\end{itemize}

This adaptive behavior allows the system to subtly flip its strategy: from maximizing hit rates when contention is low to aggressively pruning pollution as pressure mounts.  
This interplay between the replacement policy, bypassing, and runtime adaptation enables the significant performance gains observed in our \texttt{at+bypass} and \texttt{at} policies in Fig.~\ref{fig:policy_sweep_qwen}.

\subsection{A Variant of Bypassing in Inter-core Sharing Cases}
\label{BypassInterCore}

Bypassing and anti-thrashing are both designed to mitigate cache contention, but they have different capacities to handle inter-core sharing scenarios, e.g., when the workload feature a spatial group allocation dataflow (Section~\ref{ExprWorkloads}).
In an LRU cache, even if inter-core reuses are not merged in the MSHR, the cache storage can hold the data for a while, so these reuses are still likely cache hits. This property also holds true for anti-thrashing, as deprioritized data also stay in the cache for a while before getting evicted; but if some of these addresses are bypassed upon the first access, additional memory fetches will be required, taking up more memory bandwidth and leading to even poorer performance than LRU. In other words, \textbf{blindly bypassing will miss some inter-core reuse opportunities}.

Therefore a more conservative bypassing variant, which we term \texttt{gqa\_bypass}, is preferred in such cases to avoid excessive memory traffic. In our experiments, we only bypass data fetched by the slower core in a core pair (i.e. the core that commits fewer instructions) when the LLC is highly contended, while data fetched by the faster core is always cached.
This strategy, along with the anti-thrashing policy, strikes a balance between reducing cache contention and avoiding excessive memory accesses. 

\section{Cache-integrated analytical model for AI accelerators}

\label{Analytic}
\subsection{Cycle-level Simulator and Analytical Model}
Despite the relatively accurate results of cycle-level simulators, their long running time severely confines the workload size. Besides, large cases exhibit distinct computational features that cannot be simply deduced from the results of smaller ones, which we have empirically confirmed.
Hence, in addition to evaluating our design with a cycle-level simulator, we build an analytical model to extend its measurement results to real-world cases. The model does not need to precisely capture every variant of our policies; it suffices as long as it provides a proxy or bound for a properly configured policy. For instance, below we only model the optimal static bypassing that should be the upper limit of our dynamic bypassing policy.


\subsection{Overlapping Mechanism}
\label{overlapmech}
In a massively parallel architecture, a key factor that determines performance is how memory access time is hidden by computation. A conventional perspective is that GPUs can switch to other warps when one warp is blocked by ongoing memory requests. But in FlashAttention-2, all warps start with a bulk transfer from the global memory to shared memory, followed by shared memory accesses and computation.
Without explicit pipelining, it is likely that all warps in an SM are blocked by DRAM and cannot find computation instructions to hide latency. Thus, the overall time is not simply the sum or the maximum of memory transfer and computation time.

We model this behavior in our cycle-level simulator and try understanding the underlying overlapping mechanism. We notice the \textbf{time consumed by a certain type of request is determined by its corresponding bottleneck} (Eq.\ref{analyticTimeCalHit}-\ref{analyticTimeCalConflict}). For instance, cache hits and MSHR hits (memory requests merged into an existing MSHR entry) are bottlenecked by either the LLC bandwidth or the memory instruction issuing parallelism in the cores (bounded by load-store units). Similarly, MSHR misses are bottlenecked by the following with the lowest throughput: the cores, LLC, and main memory.

Inspired by works on CPU analytical models \cite{van2016analytical}, we divide cache misses into cold misses and conflict/capacity misses. Cold misses usually occur in a continuous manner, so they would easily use up MSHR resources and block the cache; in contrast, the latter are more dispersed in the instruction flow. If they are sparse enough such that MSHR resources are not exhausted (so the LLC is not stalled), these misses are in fact ``imperceptible'' to the cores since they are not blocked by the downstream memory hierarchies. In other words, conflict miss time can be hidden by computation, which translates to our final overlapping model in Eq.\ref{analyticOverallTime}.

\allowdisplaybreaks
\begin{align}
\label{analyticTimeCalHit}
    t_{hit}&=\max\left\{ \frac{n_{hit}}{N\cdot ipc_{mem}}, \frac{n_{hit}}{v_{LLC}}\right\} \\
    \label{analyticTimeCalCold}
    t_{cold}&=\max\left\{\frac{n_{cold}}{N\cdot ipc_{mem}},\frac{n_{cold}}{v_{LLC}}, \frac{n_{cold}^{'}}{bw_{cold}}\right\} \\
    \label{analyticTimeCalConflict}
    t_{cf}&=\max\left\{\frac{n_{cf}}{N\cdot ipc_{mem}}, \frac{n_{cf}}{v_{LLC}}, \frac{n_{cf}^{'}}{bw_{cf}}\right\}
\end{align}
\begin{align}
\label{analyticOverallTime}
    t=t_{hit}+t_{cold}+\max\{t_{comp},t_{cf}\}
\end{align}

Variables starting with $t$ and $n$ are the time (clock cycles) consumed by a type of request and its total number across all cores, while subscripts $cf$, $hit$, $cold$ and $comp$ stand for conflict/capacity misses, MSHR and cache hits, cold misses and computation plus local scratchpad access. $N$ is the number of cores. $n^{'}$ means the number of requests sent to the main memory after being merged by the MSHR. $ipc_{mem}$ represents the maximum possible number of global memory $\leftrightarrow$ local scratchpad transfer instructions issued each cycle in each core (similar for $ipc_{comp}$); $v_{LLC}$ is the throughput of the LLC; $bw$ is the actual main memory bandwidth (all counted in \underline{number of requests per cycle}). Also note $t_{comp}=n_{comp}/(N\cdot ipc_{comp})$. $n_{comp}$ and $n_{cold}$ can be derived from the dataflow.

\subsection{Estimating Cache and MSHR Hits}

In our dataflow assumption, each dimension can be tiled in either the spatial or temporal domain. In GQA FlashAttention operators, if the Group dimension is (at least partially) mapped to multiple cores (i.e., spatial group allocation in Section \ref{ExprWorkloads}), KV tensors will be shared among cores. This inter-core data reuse can be captured by either the MSHR or LLC: If these memory requests reach LLC closely enough, they are likely to be merged by the MSHR; otherwise, these MSHR merging events will be converted into cache hits, if they are not yet evicted. As the LLC is highly pipelined, MSHR hits and cache hits can be handled at the same throughput ($v_{LLC}$). Thus, without loss of generality, the time consumption of such inter-core reuses can be calculated by combining the reuses captured by LLC and its MSHR in a single term.

Now let us consider the effect of temporal reuses. For generic workloads, CPU analytical models like SDCM \cite{SDCM} have already proposed a mature model to predict hit rates with a given reuse distance profile. As our workload features streaming through K/V tensors repeatedly (Query and Output tensors are always bypassed from LLC as their tiles are held in the local SPM until the end of life cycle, so they are fetched from DRAM only once), the hit rate for an LRU cache is simply 100\% when the reuse distance (uniform across K/V elements) is no larger than the cache size and 0 otherwise. Our bypassing and anti-thrashing strategies attempt to make use of LLC in the latter scenario by keeping only a subset of K and V.
For ideal bypassing, this subset size should be exactly the cache size. Our self-adaptive anti-thrashing strategy is equivalent to calculating such $S_{kept}$ and the maximum possible integer $M$ that $S_{kept}=S_{work}\cdot M/2^{\text{\texttt{B\_BITS}}}\leq S_{LLC}\cdot (A-1)/A$, where $S_{kept}$, $S_{work}$ and $S_{LLC}$ correspond to the size of the prioritized subset, the working set, and LLC, respectively, while $A$ denotes associativity. Reuses of the kept subset are thus counted into $n_{hit}$. The term $(A-1)/A$ means at least one cache line per set is used to temporarily place new lines. By far, for a given dataflow, all $n$ and $n^{'}$ in Eq.\ref{analyticTimeCalHit}-\ref{analyticOverallTime} can be calculated for LRU, anti-thrashing, and ideal bypassing.

\subsection{Estimating Bandwidth Utilization}
Our experiments reveal that cold misses usually occur in bursts, so they can saturate the main memory bandwidth. Conflict and capacity misses are more dispersed in the instruction flow, so their bandwidth utilization should be calculated based on the rate of demand $v_{cf,dmd}$, which depends on $n_{cf}$, $v_{LLC}$, and the density of conflict miss instructions $\eta_{cf}$:
\begin{align}
\begin{aligned}
    v_{cf,dmd}&=\min\left\{\eta_{cf}N\cdot ipc_{mem}, v_{LLC}\right\} \\
    \eta_{cf}&=\frac{n_{cf}/ipc_{mem}}{n_{mem}/ipc_{mem}+n_{comp}/ipc_{comp}}
\end{aligned}
\end{align}
where $n_{mem}=n_{hit}+n_{cold}+n_{cf}$ is the total number of memory transfer instructions of all cores and can be calculated directly from dataflow information. As for the actual bandwidth utilization $bw_{cold}$ and $bw_{cf}$, we empirically find
\begin{align}
    bw_{cold}&\approx\theta_1 BW \\
    bw_{cf}&\approx \text{clip}(\lambda v_{cf,dmd}, \theta_2 BW, \theta_3 BW)
\end{align}
where $BW$ is the theoretical memory bandwidth calculated from the DRAM specifications, while $\theta_1$, $\theta_2$, $\theta_3$ ($\theta_2 < \theta_3$), $\lambda$ are constant coefficients for a given combination of $ipc_{mem}$, main memory, LLC bypass and replacement policy.

\subsection{Other Assumptions and Discussions}
Our analytical model does not model the effect of instruction window size, since this should not be a major bottleneck for a well-designed architecture. In our simulation, this size is selected such that it makes little difference from having an infinite one while still reasonable. We also realize that DRAM access pattern is one of the major factors that determine bandwidth utilization (i.e., $\theta_1$, $\theta_2$, $\theta_3$). We try to capture the main trend across hardware configurations and control the error to a certain limit, as predicting minor fluctuations is beyond the scope of an analytical model.

In addition, recent GPUs/NPUs differ not only in their computing and memory resources but also in several key mechanisms. The former can be modeled with different values of $n_{comp}$ and $ipc_{comp}$.
mechanisms like warp specialization and TMA in Hopper GPUs even allow $t_{cold}$ to partially overlap with $t_{comp}$, but we believe our philosophy of bottleneck analysis is applicable to all such devices.


\section{Experiments and Results}
\label{sec:exp}
\subsection{Cost Evaluation}
\label{sec:cost}
Our design is implemented in Chisel HDL and synthesized by Synopsys Design Compiler with a 15nm open-source library NanGate 15nm OCL~\cite{15nm}.
We instantiated a TMU instance with the parameters in Table~\ref{tab:sys_config}, which is synthesized at a frequency of 2.0 GHz. The report shows an area of 0.0644$mm^2$.
This area overhead is negligible compared to the typical area of a high-performance AI chip, which can be larger than 800$mm^2$, like NVIDIA H100. Since the TMU is queried at most once per fill and updated at most $0.125$ times per cycle (Section~\ref{DBPDetail}), its switching activity and energy consumption are much lower than the LLC tag/data arrays.


\vspace{-.7\baselineskip}
\begin{table}[htbp]
\centering 
\caption{Hardware Configurations}
\vspace*{-.2\baselineskip}
\label{tab:sys_config}
\begin{tabular}{c c c}
\toprule 
\textbf{Variable} & \textbf{Settings}  \\
\midrule 
LLC Slice & 32  \\
Physical Address Width & 48-bit \\
Tensor Metadata Entry Number & 8 \\
Tile Metadata Entry Number & 256 \\
Dead FIFO Depth & 16 \\

\bottomrule 
\end{tabular}
\vspace*{-1\baselineskip}
\end{table}
\vspace*{-.5\baselineskip}

\subsection{Simulation Infrastructure and Hardware Configuration}
Our strategies influence the behaviors of the memory subsystem, so a cycle-level simulator is needed for evaluation. However, current simulators fail to meet our demand for the following reasons. (1) Most of the open-source AI accelerator simulators do not support a shared LLC; (2) Existing ones like SMAUG \cite{smaug} are too slow since they involve too many details of in-core datapaths, which are unnecessary when our major concern is the LLC; (3) GPU simulators like Accel-Sim \cite{AccelSim} deeply embed GPU-specific assumptions into their code (e.g., fixed warp sizes, complex warp scheduling mechanism, running on real PTX code). These details severely confine the generality of our design and hinder efficient dataflow exploration since rewriting these kernels with manual effort is far more time-consuming and error-prone compared to directly using memory traces generated from given dataflows. 

Considering the above requirements and constraints, we adopt the cycle-level simulator proposed in LLaMCAT \cite{llamcat}, which is specifically designed for LLM workloads with flexible memory-hierarchy modeling. The simulation configurations are detailed in Table~\ref{table:simconfig}. This simulator supports shared LLC, trace-driven evaluation, and configurable dataflows, making it well-suited for the studies in this paper.

However, this simulator will still be overwhelmed when running operators from real-world LLMs. For these large-scale workloads, we have fitted an analytical model (Section \ref{Analytic}) and validated it against cycle-level simulation results. In the following sections involving cycle-level simulation, cache sizes are set to be relatively small to expose this bottleneck; they will be set to more realistic values when we scale up the problem to real-world sizes.

\begin{table}[t]
  \centering
  \caption{Simulated System Configurations}
  \vspace*{-.4\baselineskip}
  \label{table:simconfig}
  \scriptsize
\begin{tabular}{p{1.5cm}p{6.5cm}}
\hline
Basics & frequency=2.0GHz, 16 cores, 1MB-16MB LLC, 32 LLC slices \\
\hline
Core & \begin{tabular}[t]{@{}p{6cm}@{}} inst\_window\_depth=128,   num\_inst\_windows=1, 1 core=\\ 1 vector unit + private scratchpad memory, vector-len=128B \end{tabular} \\
\hline
LLC slice & associativity=8, data-latency=25, num-target=8 (per entry), mshr-num-entry=6 (per slice), alloc-on-fill, write-back, write-allocate, resp\_q\_size=64, req\_q\_size=12 \\
\hline
\begin{tabular}[t]{@{}p{1.5cm}@{}} LLC req-resp \\ arbitration \end{tabular}
  & response-queue-first \\
\hline
DRAM & DDR5\_8Gb\_x16, DDR5-3200, 4 ranks, 16 channels \\
\hline
\end{tabular}
\vspace*{-1\baselineskip}
\end{table}

\subsection{Workloads and Baselines}
\label{ExprWorkloads}
We use Gemma3-27B, Llama3-70B, and Qwen3-8B as benchmarks throughout this section. In each attention unit, these models mainly differ in the number of Q heads and KV heads. We focus on FlashAttention-2 in this paper since (1) we do not assume features like TMA and warp specialization, (2) it exhibits a more complex dataflow than GEMMs, and (3) GEMMs have been covered in the preliminary version of our paper \cite{LCM}. In addition, these models all feature grouped-query attention (GQA): it lets a group of Q heads share a single KV head to reduce KV cache size. As GQA introduces additional data sharing, its behavior may differ from classic attention.

In addition, dataflows will directly influence reusing. In GQA FlashAttention, one key factor is the mapping of the Group dimension (different Q heads within a KV head group): If it is (at least partially) assigned to different cores (\textbf{spatial group allocation}), data sharing among cores will be observed in LLC; if it is completely mapped to the temporal domain (\textbf{temporal group allocation}), there will be no data sharing among cores, resembling conventional multi-head attention.

Apart from LRU, we use DRRIP \cite{drrip} and RLR \cite{rp_bypass} as baseline replacement policies. DRRIP is implemented as a 2-bit RRIP policy with dynamic set dueling between SRRIP and BRRIP. RLR is implemented using the rule-based replacement and bypass logic described in their paper, based on per-line age and hit metadata together with a dynamically updated reuse-distance threshold.

\subsection{Efficacy of Anti-thrashing Policy}
\label{sec:anti_thrashing_efficacy}


\subsubsection{An intuitive view from a cache hit rate perspective}

\begin{figure}[t]
    \centerline{\includegraphics[width=0.98\linewidth]{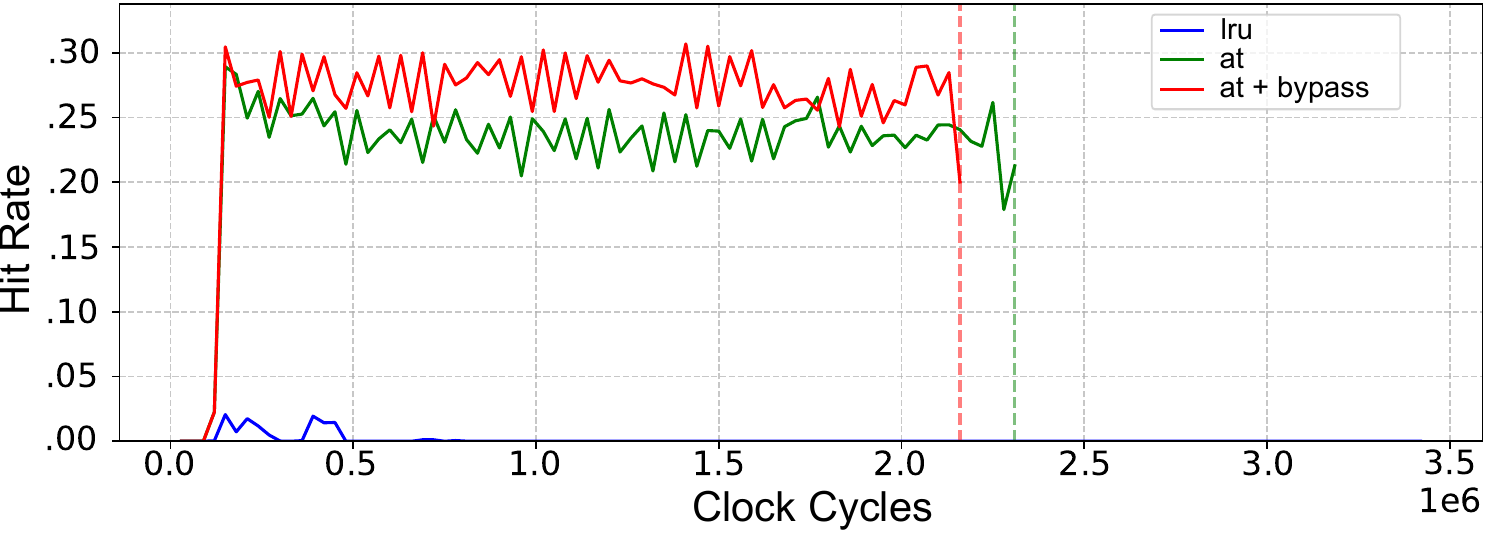}}
    \vspace*{-.8\baselineskip}
    \caption{Cache hit rate over time for LRU and our policies for a 4MB LLC (Gemma3-27B, 2K sequence length).
    }
    \vspace*{-1\baselineskip}
    \label{fig:tmu_hit_rate_comparison}
\end{figure}

Fig.~\ref{fig:tmu_hit_rate_comparison} illustrates the hit rate profile for a 4MB LLC. When thrashing occurs, LRU may exhibit a persistently low  hit rate as useful blocks are continuously evicted under long reuse distance. 
In contrast, \texttt{at} (anti-thrashing) aims to increase hit rate by keeping a fixed set of data in the cache.
The figure shows that \texttt{at} achieves a consistently higher hit rate than LRU.

\subsubsection{Comparison between anti-thrashing and LRU}


\begin{figure*}[!t]
    \centerline{\includegraphics[width=0.6\linewidth]{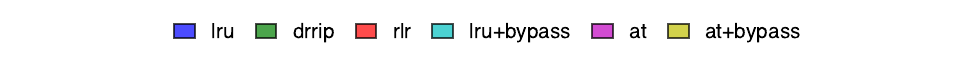}}
    \vspace*{-0.95\baselineskip}

    \centering

    \subfloat[\scriptsize Gemma3-27B, temporal group allocation, 2K sequence length]{\includegraphics[width=0.5\linewidth]{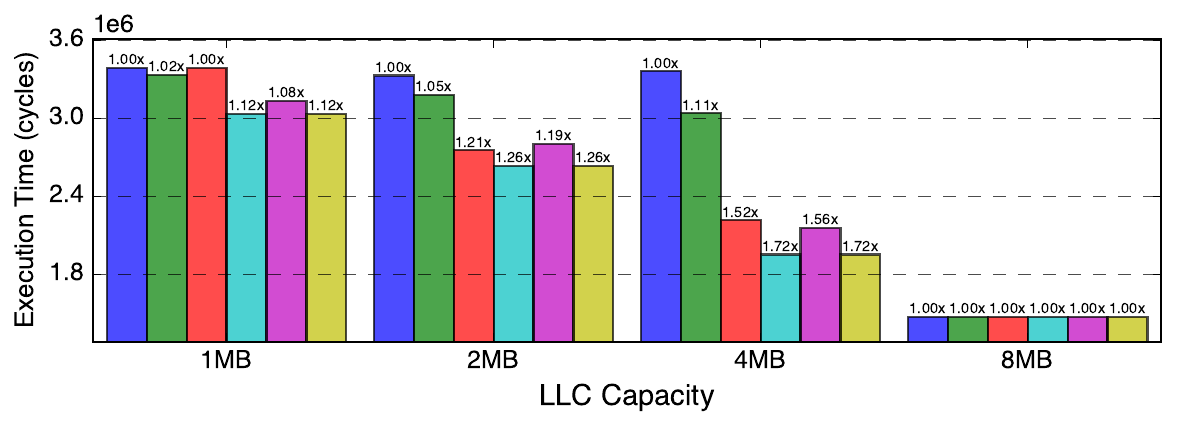}}
    \subfloat[\scriptsize Gemma3-27B, temporal group allocation, 4K sequence length]{\includegraphics[width=0.5\linewidth]{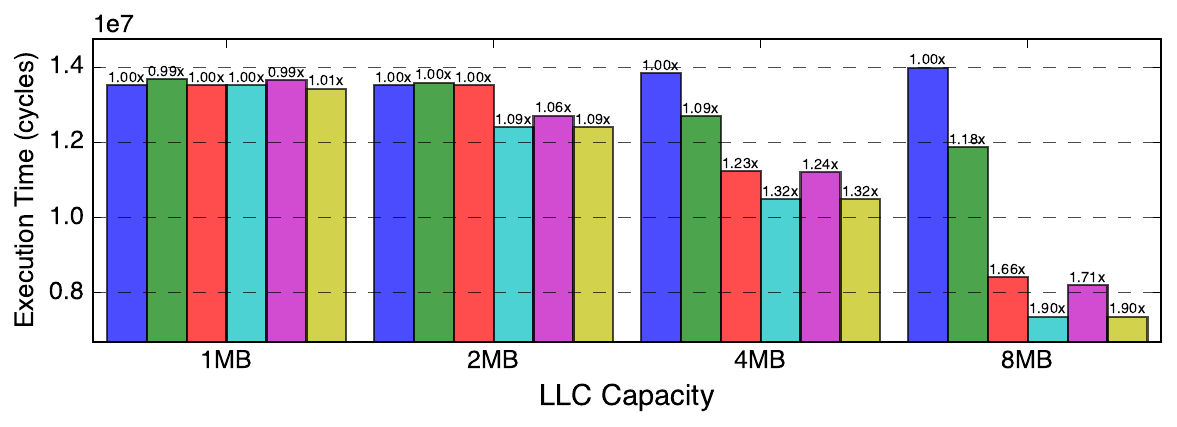}}

    \vspace*{-0.25\baselineskip}
    \subfloat[\scriptsize Qwen3-8B, spatial group allocation, 2K sequence length]{\includegraphics[width=0.5\linewidth]{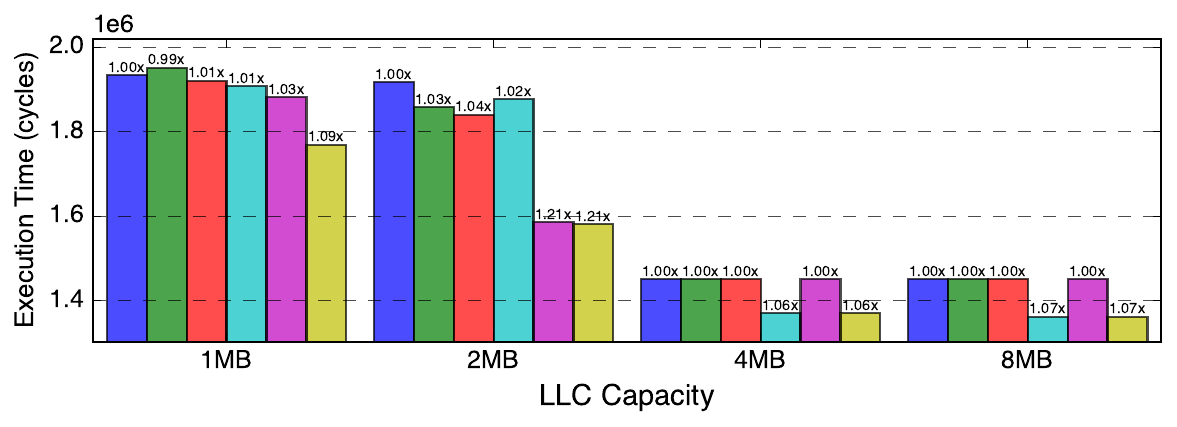}}
    \subfloat[\scriptsize Qwen3-8B, spatial group allocation, 4K sequence length]{\includegraphics[width=0.5\linewidth]{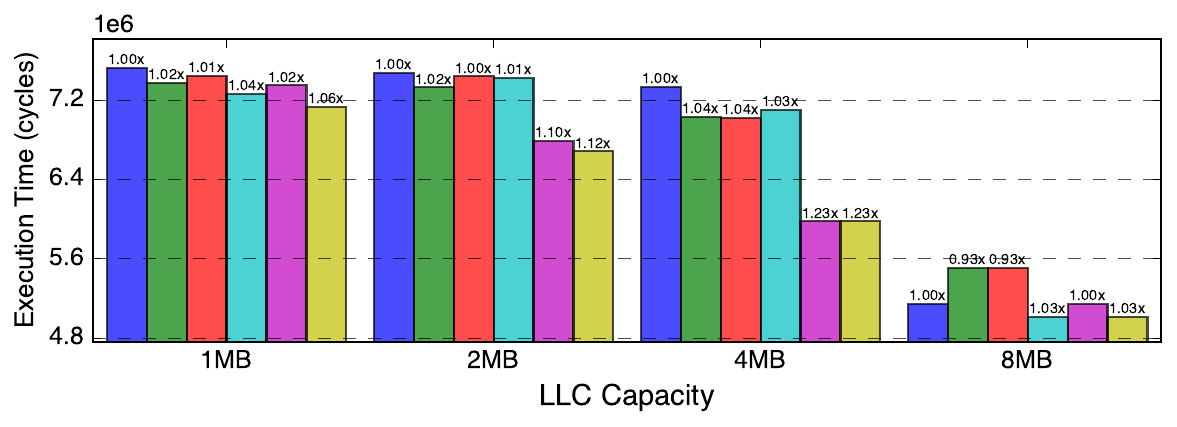}}

    \vspace*{-0.25\baselineskip}
    \subfloat[\scriptsize Llama3-70B, spatial group allocation, 2K sequence length]{\includegraphics[width=0.5\linewidth]{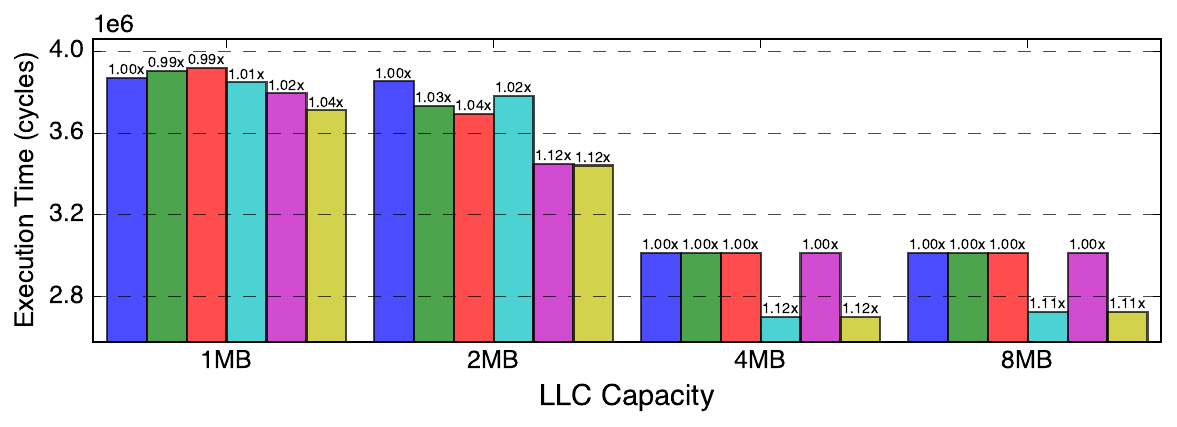}}
    \subfloat[\scriptsize Llama3-70B, spatial group allocation, 4K sequence length]{\includegraphics[width=0.5\linewidth]{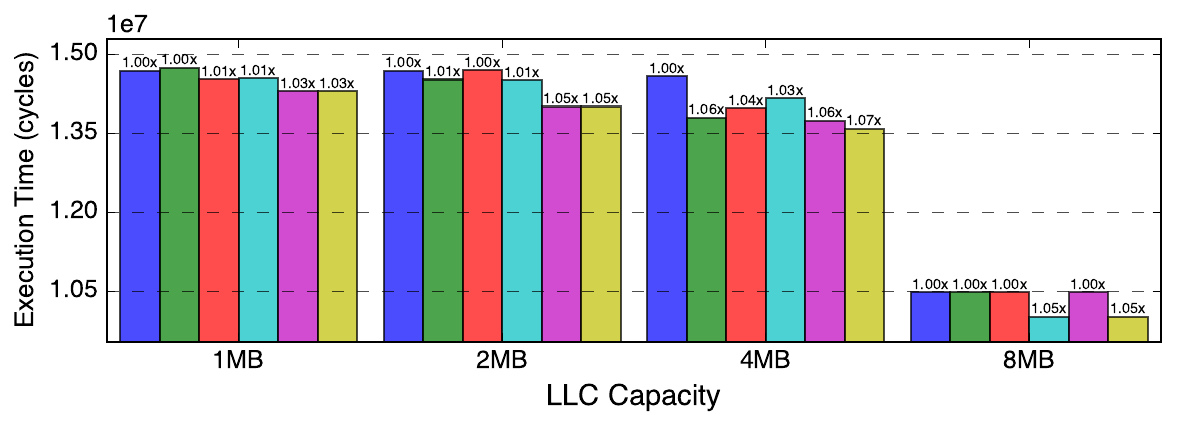}}

    \caption{Execution time for different LLC replacement policies and capacities. \texttt{at}: anti-thrashing. Unless otherwise specified, \texttt{bypass} for spatial group allocation cases refers to the \texttt{gqa\_bypass} variant.}
    \vspace*{-1\baselineskip}
    \label{fig:policy_sweep_qwen}
\end{figure*}

Fig.~\ref{fig:policy_sweep_qwen} presents the execution time (lower is better) of \texttt{lru}, \texttt{at} and several baseline policies across various LLC capacities in Gemma and Qwen cases. 
The figure also evaluates bypassing strategies for a broader context, though they will be discussed separately in Section~\ref{sec:dynamic_cache_bypassing}.
With moderate LLC capacities (2MB and 4MB), \texttt{at} consistently outperforms the standard LRU approach. With a 4MB LLC, it achieves a $1.56\times$ speedup in the Gemma 2K case and a $1.23\times$ speedup in Qwen 4K. Similar improvements are observed with a 2MB LLC, indicating that the scoring mechanism effectively prioritizes specific cache lines, thereby reducing cache thrashing.

When the LLC capacity is much smaller than the working set, such as 1MB, the performance of each policy degrades significantly due to severe thrashing. However, \texttt{at} still manages to provide a modest $1.08\times$ speedup in the Gemma 2K case and $1.03\times$ in Qwen 2K. This suggests that while the scoring mechanism can help mitigate some thrashing, its effectiveness is still limited under extremely constrained conditions.

As the LLC capacity increases to 8MB, which is exactly the size of the active working set of the Gemma3-27B 2K case, both policies perform similarly with negligible difference. This is expected as LLC can now hold the entire working set, so the impact of replacement policies is minimized.


\begin{figure}[t]
    
    \centering
    \vspace*{-0.6\baselineskip}
    \subfloat[\scriptsize Gemma3-27B, temporal group allocation, 4K sequence length]{\includegraphics[width=1\linewidth]{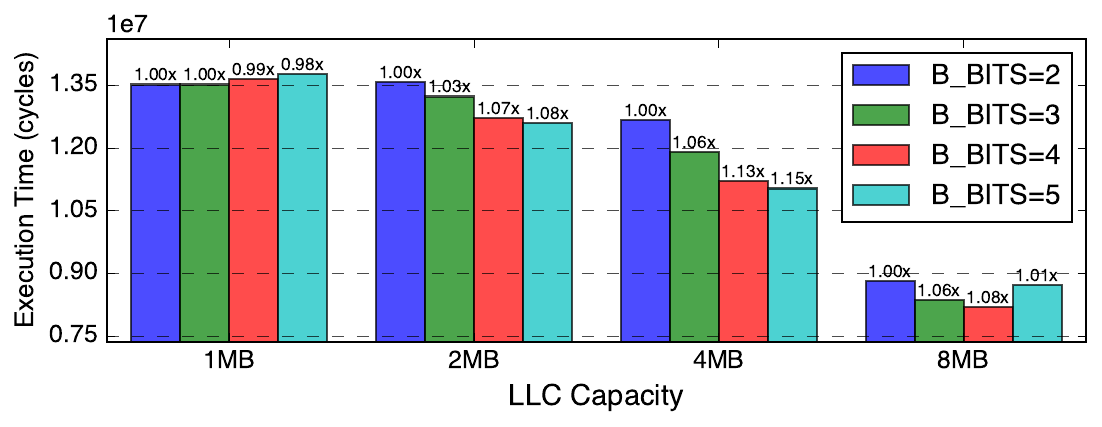}}

    \vspace*{-0.6\baselineskip}
    \subfloat[\scriptsize Qwen3-8B, spatial group allocation, 4K sequence length]{\includegraphics[width=1\linewidth]{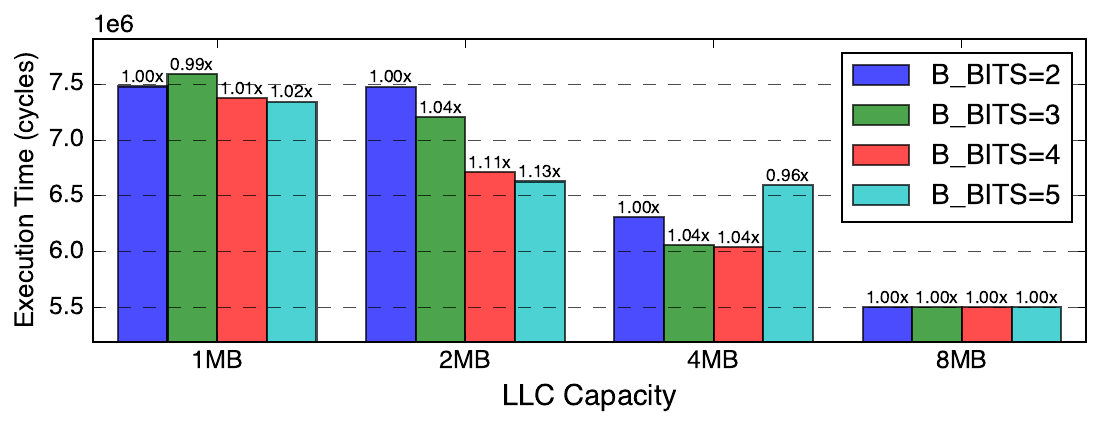}}
    \caption{Execution time for different anti-thrashing configurations (\texttt{B\_BITS}) and capacities. Lower values indicate better performance.}
    \vspace*{-1\baselineskip}
    \label{fig:tmu_bits}
\end{figure}

\begin{figure}[t]
    
    \centering
    \vspace*{-0.6\baselineskip}
    \subfloat[\scriptsize Gemma3-27B, temporal group allocation, 4K sequence length]{\includegraphics[width=1\linewidth]{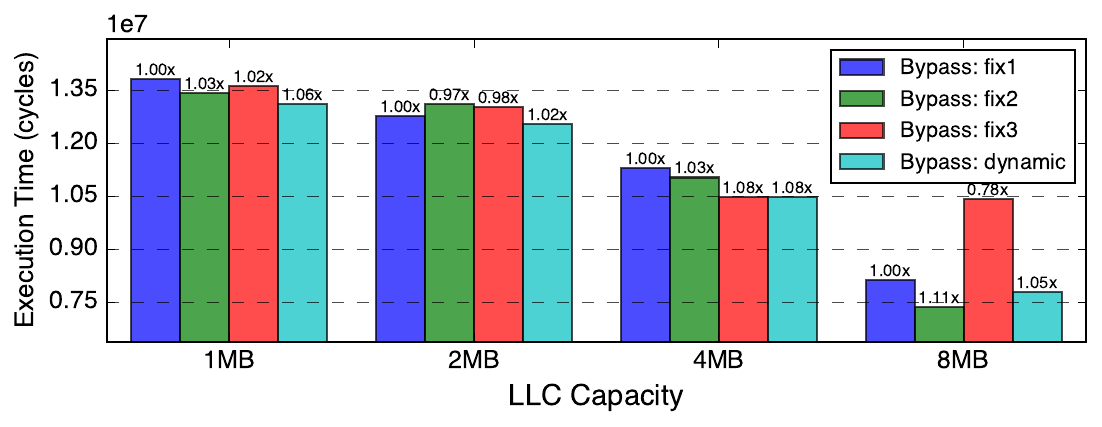}}

    \vspace*{-0.6\baselineskip}
    \subfloat[\scriptsize Qwen3-8B, spatial group allocation, 4K sequence length]{\includegraphics[width=1\linewidth]{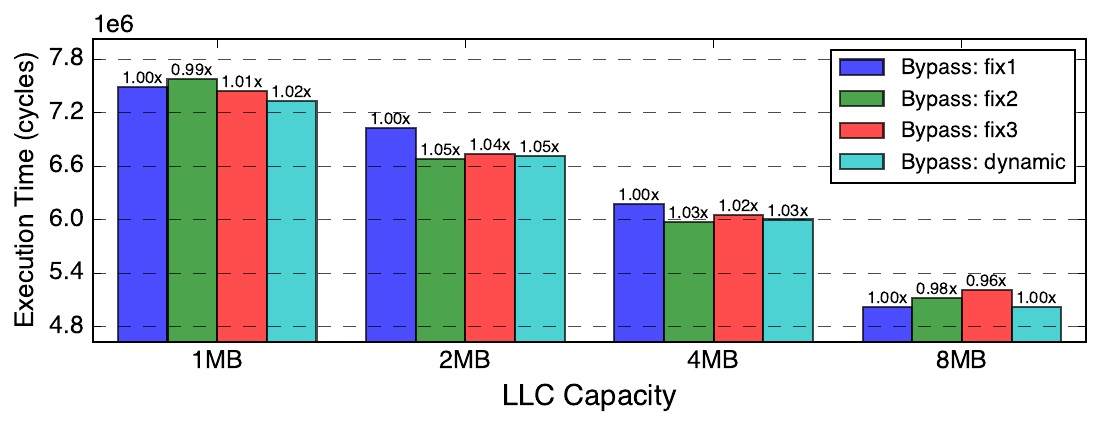}}
    \caption{Execution time for different cache bypass policies, normalized against \texttt{fix1}. \texttt{at} is always enabled to show the additional benefits of bypassing.}
    \vspace*{-1\baselineskip}
    \label{fig:tmu_bypass_comparison}
\end{figure}


\subsubsection{Design choices of anti-thrashing policy}
\label{design_choices}
A critical design parameter for the anti-thrashing policy is the number of tag bits used for scoring (\texttt{B\_BITS}). In the evaluated setting, 4 bits are used, implying 16 distinct priority levels for cache lines. The choice of the number of bits involves a trade-off:
\begin{itemize}
    \item \textbf{More bits} (e.g., 3-4 bits, yielding 8-16 levels) 
    allows for a finer-grained classification of cache lines, which is particularly advantageous for moderately sized caches. It enables more nuanced prioritization without excessively restricting the working set available to the underlying LRU component, which manages lines with the same TMU score.
    However, if the cache is large enough that simple LRU already performs well, the overhead of managing many fine-grained levels might not provide additional benefits.
    \item \textbf{Fewer bits} (e.g., 1-2 bits, yielding 2-4 levels) results in a coarser classification. 
    It is an effective strategy for aggressive anti-thrashing policy in very small or highly contended caches, as it partitions data into fewer distinct priority groups. However, for larger caches where thrashing is less of an issue, its restrictive nature could make the policy degenerate towards standard LRU, resulting in the premature eviction of useful data.
\end{itemize}
The optimal number of bits is likely application-specific and correlated with the cache size, associativity, and typical data access patterns. Our selection of 4 bits strikes a balance, providing distinct priority levels without excessive complexity. Fig.~\ref{fig:tmu_bits} illustrates the performance with varying \texttt{B\_BITS} and cache sizes. While there is no definitive optimal configuration across all scenarios, 4 bits consistently exhibit stable speedups.

\subsection{Efficacy of Dynamic Cache Bypassing}
\label{sec:dynamic_cache_bypassing}
Since \texttt{at} already produces a stable improvement under both types of dataflows, here we focus on the \textbf{additional} effects of bypassing by always enabling anti-thrashing.
We first compare our dynamic bypassing policy against several static ones with fixed bypassing gears (\texttt{B\_GEAR}) to show its flexibility and near-optimality, followed by discussions on its performance under different dataflows. Finally we compare bypassing with anti-thrashing to better understand their own advantageous regions and how they can collaborate to offer a higher speedup.

\subsubsection{Comparison between dynamic and static bypassing}

\begin{figure}[t]
    \centering 
    \subfloat[\scriptsize Gemma3-27B, temporal group allocation, 2K, 2MB LLC]{\includegraphics[width=1\linewidth]{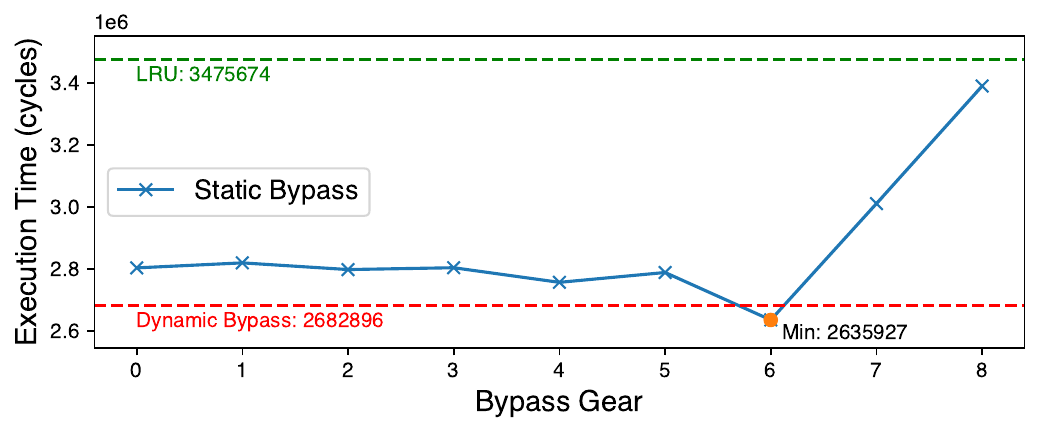}} \\
    \vspace{-.6\baselineskip}

    \subfloat[\scriptsize Qwen3-8B, spatial group allocation, 2K sequence, 1MB LLC]{\includegraphics[width=1\linewidth]{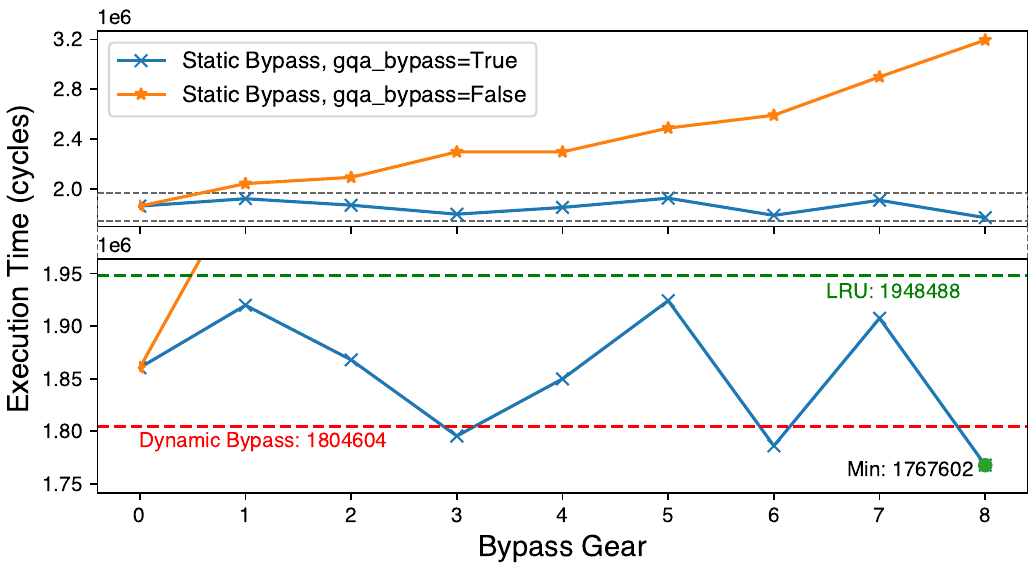}} \\
   
    \caption{Execution time of static and dynamic bypassing with varying \texttt{B\_GEAR} values. In both cases, bypassing uses 3 bits, so $0\leq$ \texttt{B\_GEAR} $\leq 8$. \texttt{at} is always enabled, so \texttt{B\_GEAR} $=0$ degenerates to ordinary \texttt{at}.}
    \vspace*{-1\baselineskip}
    \label{fig:StaticVSDynamic}
\end{figure}


We first compare our dynamic bypass policy against three static ones (\texttt{fix1}, \texttt{fix2}, \texttt{fix3}), which employ fixed \texttt{B\_GEAR}s in an ascending order. Fig.~\ref{fig:tmu_bypass_comparison} presents the execution times for each policy across LLC capacities from 1MB to 8MB.

The results highlight the fundamental drawback of static approaches: no single policy is optimal across all scenarios. The \texttt{fix3} policy, the most aggressive among them, performs the best when the cache is highly contended. However, this same aggressiveness becomes detrimental as cache capacity increases, making it the worst-performing policy at 8MB. Conversely, a less aggressive policy is more suitable for larger caches but is suboptimal under high contention.

Our dynamic policy overcomes this limitation by adapting its strategy under the guidance of eviction rate. It achieves the shortest execution time for 1MB and 2MB capacities and delivers performance close to the best static policy (\texttt{fix3}) at 4MB. Fig. \ref{fig:StaticVSDynamic} further dives into 2 specific settings, one for each dataflow, and compares our dynamic policy with static ones of all possible gears. Under both dataflows, our policy can achieve near-optimality (within 3\% difference), thereby demonstrating its robustness.

\subsubsection{Bypassing under different dataflows}
\label{BypassingSpatial}




As explained in Section \ref{BypassInterCore}, bypassing can miss some inter-core reuses.
The \texttt{gqa\_bypass} variant is thus designed for such cases. Unless otherwise specified, \texttt{bypass} used in spatial group allocation dataflows in all the experiments refer to this variant. Fig. \ref{fig:StaticVSDynamic}(b) also includes an ablation study to show its efficacy. Without \texttt{gqa\_bypass}, the performance of static bypassing (orange line) quickly degrades below LRU: the more it bypasses, the worse it performs. The only exception is at \texttt{B\_GEAR} $=0$, which means ``to bypass nothing'' and degenerates to the ordinary anti-thrashing policy (remember \texttt{at} is always enabled in this subsection). A dynamic bypassing policy in this case will finally find its performance within the range of the orange line, which is not likely to be better than LRU. However when \texttt{gqa\_bypass} is enabled, bypassing can outperform LRU (blue line and red dashed line), despite the modest amount. 




\subsubsection{Connection between bypassing and anti-thrashing}
\label{BypassAndAT}
After analyzing the properties of our bypassing policy, here we discuss how it interacts with anti-thrashing. In Fig. \ref{fig:policy_sweep_qwen}(a) and (b), \texttt{lru+bypass} performs better than \texttt{at}, because the former can make use of the whole cache to pin data, while \texttt{at} has to set aside at least one cache line per set to temporarily store the victims. But in (c) and (d), the trend differs since inter-core reuse exists and we have to adopt the more conservative \texttt{gqa\_bypass} variant. However, in all these cases, combining \texttt{at} and \texttt{bypass} does produce the best speedups. One reason is that anti-thrashing can further filter out over-sized working sets when bypassing alone is not precise enough, for example, when it is still under dynamic adaption. It is also worth noting that when cache sizes are sufficient, bypassing can still bring additional speedups, e.g., $1.07\times$ in Fig.~\ref{fig:policy_sweep_qwen}(c) 8MB. This is because bypassing in this case reduces cache bandwidth pressure and leverages the idle DRAM bandwidth. Later in Section \ref{subsec:analytical_eval}, with the help of the analytical model, we will explain this also helps \texttt{at+bypass} to outperform \texttt{at} in some other cases, e.g., Fig. \ref{fig:policy_sweep_qwen}(c) 1MB.

\subsection{Efficacy of Dead Block Prediction}
\label{sec:dbp_efficacy}

DBP is designed to enhance cache efficiency by proactively identifying and evicting data that is no longer needed. Its effectiveness is particularly pronounced in workloads with distinct temporal phases, such as when multiple KV heads are mapped to the temporal domain, and in the multi-batch inference scenario evaluated here. 
In this context, data from completed batches becomes ``dead'' and pollutes the cache, contending for resources with active data from the current batch. To precisely quantify the benefits of DBP, we compare the performance of DBP---in cooperation with the anti-thrashing and bypassing mechanisms---against the standard \texttt{at+bypass} policy.

\begin{figure}[t]
    \centering
    \includegraphics[width=\linewidth]{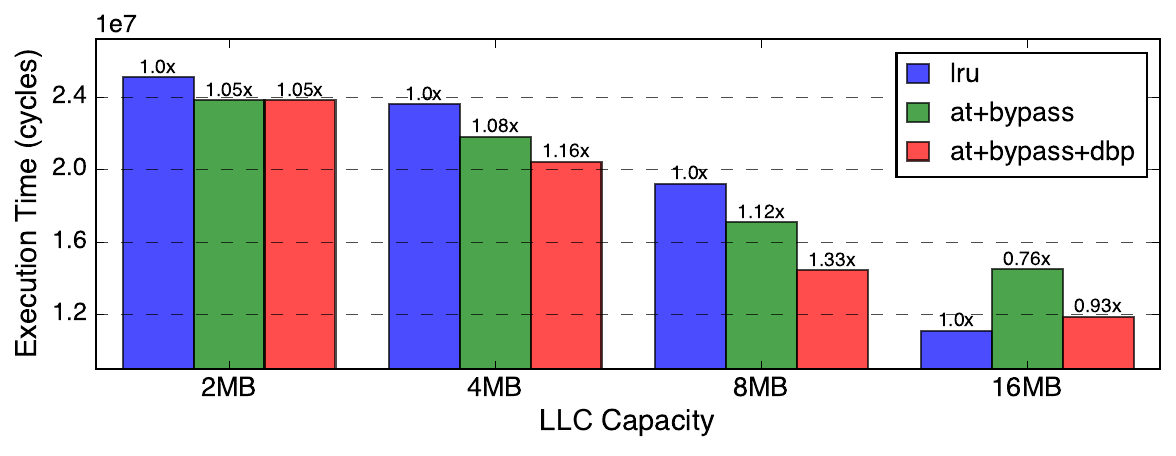}
    \vspace*{-1.8\baselineskip}
    \caption{Performance impact of DBP in a multi-batch scenario (Gemma3-27B, 4K sequence length, 2 batches). The chart compares execution time for \texttt{at+bypass} (anti-thrashing and bypassing) against the one which adds DBP.
    }
    \vspace*{-1\baselineskip}
    \label{fig:dbp_efficacy_comparison}
\end{figure}

The results in Fig.~\ref{fig:dbp_efficacy_comparison} show that DBP offers significantly better performance, particularly when the cache is moderately sized. With a 4MB LLC, the addition of DBP provides a $\mathbf{1.07\times}$ speedup. Similarly, for an 8MB LLC, DBP provides a strong $\mathbf{1.19\times}$ speedup. 
In these settings, data from the current and previous batches contend for the limited cache space. Standard \texttt{at} cannot directly distinguish useful data from the current batch and obsolete data from previous batches. However, DBP resolves this by identifying obsolete data as dead and thereby clearing useless data.



The situation is different when the cache size is extremely constrained, such as 2MB, which is much smaller than the working set size.
Here, the cache is so small that \texttt{at} already evicts out most of the incoming lines, including those dead, which explains the marginal improvement brought by DBP ($1.01\times$).
However, this is a special case; under normal circumstances, dead blocks must still be explicitly distinguished to provide accurate information for cache replacement.

Another scenario is that LRU provides the best performance at 16MB, where the entire working set fits into the cache without causing eviction. LRU is able to switch to another working set due to its nature of retaining recently used data, and the overhead of DBP in tracking dead blocks becomes unnecessary.
Policy \texttt{at}, however, performs the worst in this case, as it fails to utilize the full cache space due to the pollution of dead blocks.

In conclusion, DBP is particularly powerful for workloads with periodical data retirement under moderately sized caches.
It works in harmony with other cache management strategies so that they can filter out the data to be held in the cache using their own criteria without interference from dead blocks.

\subsection{Comparison with Baselines and Energy Cost Evaluation}
\begin{figure}[t]
    \centering
    \includegraphics[width=\linewidth]{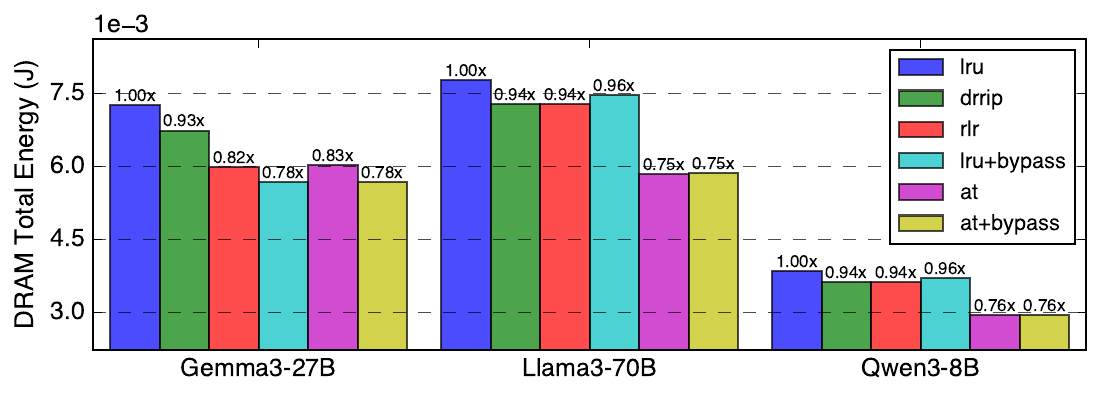}
    \vspace*{-1.8\baselineskip}
    \caption{DRAM energy consumption of various models and LLC policies under a typical setting (2MB LLC, 2K sequence length). Numbers on top of each bar shows the DRAM energy consumed relative to LRU.
    }
    \vspace*{-1\baselineskip}
    \label{fig:DRAMEnergy}
\end{figure}

In addition to LRU, Fig. \ref{fig:policy_sweep_qwen} also shows the execution time of two baselines: DRRIP and RLR. They both outperform LRU under moderate cache sizes because they can adapt to thrashing conditions to some extent, with typical examples like 2MB and 4MB in Fig. \ref{fig:policy_sweep_qwen}, although in other cases where cache storage pressure is small, they can be overly conservative, like Fig. \ref{fig:policy_sweep_qwen}(d) @8MB. But overall, they still cannot surpass our metadata-aided \texttt{at+bypass} under all cases. We further compare with two conventional miss-triggered LLC prefetchers: an adjacent-line prefetcher and an stride prefetcher. Across all configurations their speedup over LRU remains within $0.95-1.03\times$, despite a prefetch accuracy of $0.94-1.00$. Prefetching hides memory latency, but this workload is already bound by DRAM bandwidth. We have also measured the DRAM energy consumption across various LLMs and LLC policies, using DRAMPower 5.0 \cite{DRAMPower}, as shown in Fig. \ref{fig:DRAMEnergy}. Overall, the figure aligns with the execution time results in Fig.~\ref{fig:policy_sweep_qwen}, with our policy \texttt{at+bypass} featuring the lowest energy consumption ($0.75-0.78\times$ of LRU) among all policies (e.g., RLR: $0.82-0.94\times$). This is mainly attributed to the reduced execution time and DRAM accesses.

\subsection{Evaluation on Longer Contexts via Analytical Model}
\label{subsec:analytical_eval}

To verify the effectiveness of our design in real-world long-context scenarios and explore future-scale design choices, we turn to the analytical model in Section~\ref{Analytic}. We restate that the model is intended to fit the performance for several key variants of our policies instead of every possible configuration.

\subsubsection{Validating the analytical model}
Before large-scale experiments, we need to verify the accuracy of the analytical model first. We perform validation between the cycle accurate simulator and the analytical model across the following design and mapping space: 
\begin{itemize}
    \item LLC size: 1MB, 2MB, 4MB, 8MB
    \item policies: \texttt{lru}, \texttt{dbp}, \texttt{at+dbp}, \texttt{bypass+dbp}, \texttt{all}
    \item bypass variants: \texttt{fix1}, \texttt{fix3}, \texttt{optimal}
    \item \texttt{gqa\_bypass}: true if spatial group allocation else false
    \item workloads: Gemma3-27B, Llama3-70B, Qwen3-8B
    \item sequence length: 2K, 4K, 8K
    \item dataflow: \{spatial/temporal\} group allocation
\end{itemize}

\begin{figure}[t]
    \centering
    \vspace*{-.2\baselineskip}\includegraphics[width=\linewidth]{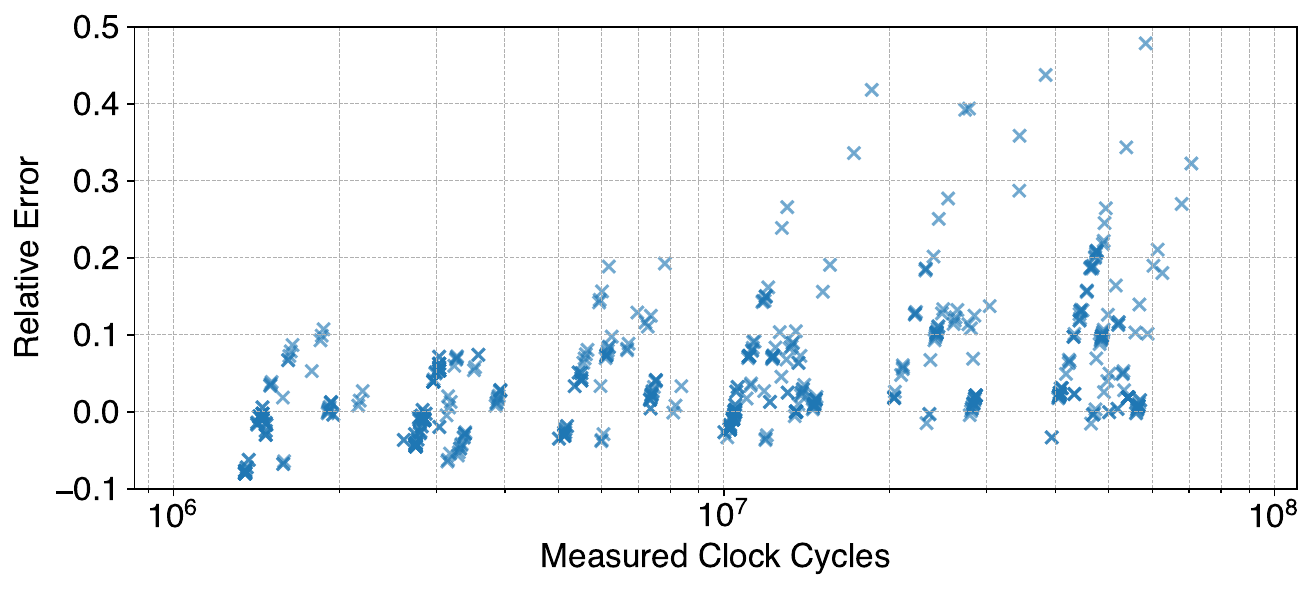}
    \vspace*{-1.8\baselineskip}
    \caption{Comparison between simulation and analytical model results ($R^2=0.997$, Kendall $\tau=0.925$)}
    \vspace*{-1\baselineskip}
    \label{fig:validateanalytical}
\end{figure}

In the listing above, \texttt{all=at+bypass+dbp}; we always enable \texttt{dbp} except for \texttt{lru} since it provides a clean separation between adjacent working sets, making it easier for \texttt{at} and \texttt{bypass} to perform well. \texttt{optimal} is a static bypassing variant with optimal \texttt{B\_GEAR}, which is the upper limit of our final \texttt{dynamic} bypassing policy. We use it as a delegate since we have empirically confirmed the near-optimality of our \texttt{dynamic} policy in Section \ref{sec:dynamic_cache_bypassing}. As for the dataflow, we always map the Group dimension in the temporal domain inside other embarrassingly parallel dimensions such as batch, K/V heads, and the sequence length of the Q tensor. We do not include other design space parameters like the number of cores, SIMD lanes, and LSUs in this section for the following reasons: (1) We are extending the measurement results for the same hardware setting as in previous sections; (2) Changing the cache size is already sufficient to sweep through the spectrum from compute-bound to memory-bound cases. A total of 648 data points ($=4$ (cache\_size) $\times 3$ (seq\_len) $\times 6$ (workload\&dataflow) $\times 9$ (policies: $3$ (\texttt{lru}, \texttt{dbp}, \texttt{at+dbp}) $+3$ (bypass variant) $\times 2$ (\texttt{bypass+dbp}, \texttt{all}))) have been extracted from the design space, of which the (measured execution time, reletive error) pairs are plotted in Fig. \ref{fig:validateanalytical}(a). We deem this scale to be sufficient since the execution time has spanned almost 2 orders of magnitude. The samples show a mean absolute percentage error of 5.68\%, while the maximum observed percentage error is 48.9\%. Despite the large error in sporadic cases, the overall accuracy is already acceptable for an analytical model. Moreover, the fitting result shows a correlation coefficient of $R^2=0.997$ and a Kendall $\tau=0.925$, which indicates that our model preserves order to a large extent. This can be a more important property when applied to design space exploration problems.


   

\subsubsection{Evaluating larger workloads}

\begin{figure}[t]
  \centering
  \includegraphics[width=1.0\linewidth]{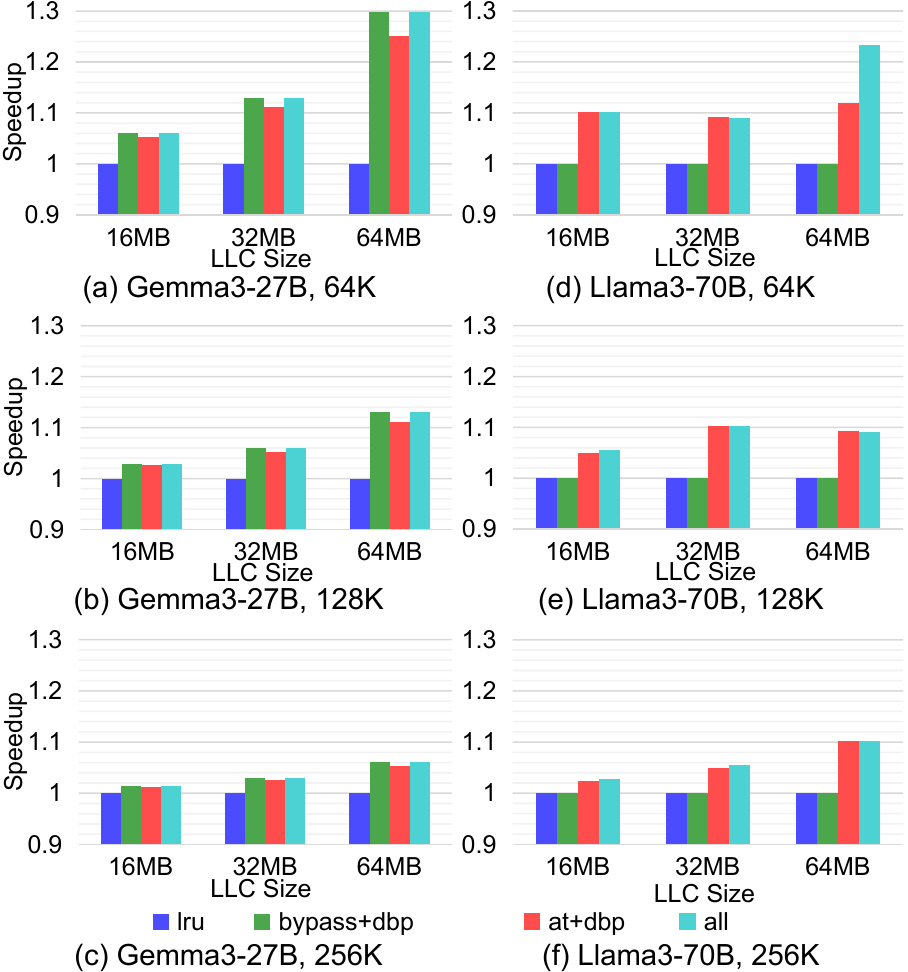}
  \vspace*{-1.3\baselineskip}
  \caption{Speedups (normalized to LRU) of the proposed cache orchestration strategies under longer sequence lengths, evaluated with the analytical model.}
  \label{fig:analytical_results}
  \vspace*{-1\baselineskip}
\end{figure}

The workloads include Gemma3-27B, Llama3-70B, and Qwen3-8B, with sequence lengths extending to 64K, 128K, and 256K. To align with previous sections, here we also apply temporal group allocation to Gemma3 and spatial to other cases. The LLC size is swept across 16MB, 32MB, and 64MB.

Fig.~\ref{fig:analytical_results} illustrates the speedup of different configurations: the full DCO design (\texttt{all}), \texttt{optimal} bypassing with dead block prediction (\texttt{bypass+dbp}), anti-thrashing with dead block prediction (\texttt{at+dbp}), and baseline LRU.

\textbf{Baseline Analysis (LRU):} We find the absolute execution time for LRU maintains almost identical across 16MB, 32MB, and 64MB caches. This is because the working set size is far beyond the LLC size, leading to severe thrashing and a very low hit rate that is irrelevant to the cache size. All speedups reported in Fig.~\ref{fig:analytical_results} are normalized to this baseline.

\textbf{Performance on Gemma3:} For Gemma3-27B (a)-(c), our DCO design achieves significant speedups, peaking at $1.30\times$ with a 64MB cache. We observe that the efficacy of the combined strategy scales positively with cache size. 
Similar to previous simulation results, \texttt{bypass+dbp} also outperforms \texttt{at+dbp} since the latter cannot use the whole cache to pin data. However, adding \texttt{at} to \texttt{bypass+dbp} (\texttt{all}) here only makes marginal difference. This is because we are assuming \texttt{optimal} bypassing and neglecting non-ideal factors in \texttt{dynamic} bypassing such as the period of dynamic adaption.

\textbf{Performance on Llama3 and Generalization:} The results for Llama3-70B (d)-(f) exhibit a different characteristic. Here, \texttt{bypass+dbp} yields negligible speedup ($\approx 1.0$), indicating that without \texttt{at}, the \texttt{gqa\_bypass} variant \textbf{alone} struggles to improve over LRU in cases with inter-core reuse. \texttt{at+dbp}, as expected, remains highly effective and contributes to the majority of the performance gains (up to $1.12\times$). Surprisingly, adding \texttt{bypass} to \texttt{at+dbp} still sometimes brings additional benefits, e.g., Llama3-70B 64K@64MB, 128K@16MB, 256K@16MB and 32MB. After dissecting the running time, we discover that bypassing can relieve the burden on cache bandwidth in these cases, while in previous sections we only see similar phenomena when LLC is large enough to hold the entire working set. This again demonstrates that our measures take effect from different perspectives and can collaborate to maximize speedup.
We further validate Qwen3-8B using the same method and found its performance trends similar to those of Llama3-70B in Fig.~\ref{fig:analytical_results}(d)-(f).

\subsection{Cache vs. SPM under Mixed-precision Quantization Cases}
\begin{figure}[t]
  \centering
  \includegraphics[width=1.0\linewidth]{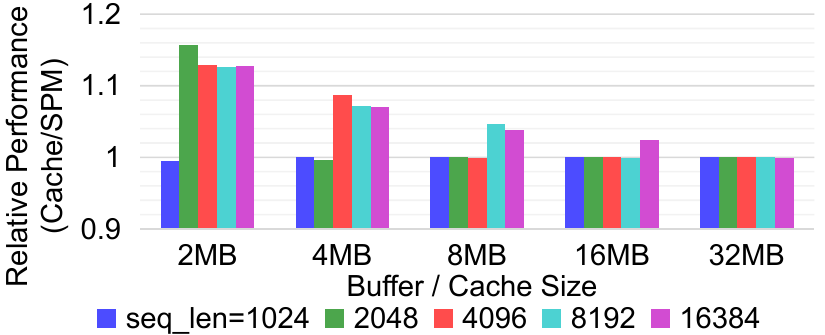}
  \vspace*{-1.3\baselineskip}
  \caption{Speedups of shared cache over SPM for MPMRF logit in Qwen3-8B}
  \label{fig:CacheVsSPMDCO}
  \vspace*{-1\baselineskip}
\end{figure}

Although SPMs introduce less hardware overhead and can be more performant for regular workloads, caches are more flexible in dynamic and data-dependent cases. As an example, we adopt the same MPMRF way to compute attention as our preliminary work \cite{LCM} and compare the execution time of systems with a shared cache and those with a shared SPM using the analytical model (Fig. \ref{fig:CacheVsSPMDCO}). We refer readers to our preliminary paper \cite{LCM} for details of MPMRF. The reuse distance profile is sampled using real Qwen3-8B weights. Results show that (1) caches can outperform SPMs, and (2) under shorter sequence lengths, there would be larger performance gaps (up to $1.16\times$).
However, this projected margin is smaller than the worst-case error of our analytical model, so it must not be read as a numerically confirmed speedup.
This is because (1) a cache can capture fine-grained data reuse automatically: even when the SRAM cannot accommodate a whole K head, a cache can flexibly reuse some of the embedding vectors between rounds of accesses, while an SPM has to repetitively fetch the same tensor, and (2) the larger a K head becomes, the less data reuse a cache can capture. If a K head grows so large that the minimum reuse distance exceeds the cache capacity, cache performance will drop to the level of an SPM.

\section{Related Work}
Previous research on optimizing GPUs and multi-core AI accelerators can be categorized into two areas: chips with global on-chip buffers and those without. Furthermore, many studies focus on cache management across CPUs, GPUs, and embedded processors. Our work significantly diverges from them, particularly in the target architecture and workloads.

\subsection{Global Buffer Management in Multi-core AI Accelerators}
Pin-or-Fuse \cite{PorF} and OnSRAM \cite{onsram} enhance on-chip data reuse using software compilation optimizations. However, as Telamalloc \cite{tela} emphasizes, compilation is a crucial step in launching applications, which can be slowed down by these additional management passes. In contrast, our method with a global cache does not require complex buffer management and substantially reduces compilation efforts.

\subsection{Optimizing Multi-core AI Accelerators w/o a Global Buffer}
Not all works from academia include a global buffer in their design. For example, Tangram \cite{TANGRAM} and Atomic Dataflow \cite{ATOM} instead create specific on-chip data transfer routes to facilitate data reuse. Although these designs may lower resource demand on chips, they are often tailored for certain applications, primarily convolutional neural networks (CNN). 

Nonetheless, using a shared buffer effectively reduces the bandwidth gap between private buffers and DRAMs. Besides, accelerators with a shared buffer, in contrast to buffer-free ones, do not rely on any assumptions about the graph-level structure of the workload, making them suitable for a broader range of applications and deserving increased focus.

\subsection{Cache Management and Bypassing}
Various strategies \cite{MPPP, DBP1, DBP2} forecast inactive blocks in CPU caches, enabling timely dead block clearance or bypassing the last access. They aim to increase prediction accuracy and reduce hardware costs using sampling- and counter-based methods. By contrast, our focus is on application-specific platforms that provide detailed insights into memory access patterns, eliminating the need for general-purpose techniques. 

Recent works use learning-based methods to optimize cache bypassing. RLR \cite{rp_bypass} uses reinforcement learning as an offline tool to design efficient replacement and bypassing policies for CPU caches, which differs from our primary concern. \cite{dbypass} introduces the concept of dynamic cache bypassing, which is insightful for our design. It employs an online learning mechanism to determine, on-the-fly, the optimal proportion of thread blocks that should bypass the cache for these “medium locality” accesses. However, it only targets GPUs and does not explore the coordination between bypassing and replacement policy, which is a unique perspective that our work features.



\subsection{AI accelerator analytical models}
Timeloop \cite{timeloop} is a flexible modeling platform for DNN accelerators, delivering performance and energy estimation to evaluate trade-offs. Stream \cite{stream} facilitates layer fusion and supports various multi-core architectures for comprehensive DSE. The tool in \cite{TANGRAM} allows for energy-efficient dataflow and mapping exploration, utilizing a tiled multi-core architecture where each core mimics an Eyeriss-style \cite{chen16} engine. FRAME \cite{frame} uses a roofline model to assess different operators across various NNs, providing insights into the compute/memory ratio in accelerator design. Nonetheless, these models often exclude caches in hardware setups. Our cache-integrated model opens up new possibilities for DSE problems.
\section{Conclusion}
This work introduces DCO, a hybrid architecture that optimizes data reuse in multi-core AI systems with a shared cache. Results show dead block prediction, anti-thrashing, and bypassing have their own advantageous ranges and can be combined to achieve higher speedup (up to $1.9\times$).
We also build and validate an analytical model to evaluate our design under practical and larger workloads. Evaluation via this model further confirms our policies outperform conventional LRU caches under longer sequence lengths, though moderate cache sizes are still preferred:
going to either extreme will make the replacement/bypass policy less effective.
\section*{Acknowledgment}
This work was supported by ZTE Industry-University-Institute Cooperation Funds under Grant No. IA20250923007.

This work was also partially
funded by Hong Kong RGC GRF16219825.

\bibliographystyle{IEEEtranS}
\bibliography{sample-base}

\newpage

 

\begin{IEEEbiography}[{\includegraphics[width=1in,height=1.25in,clip,keepaspectratio]{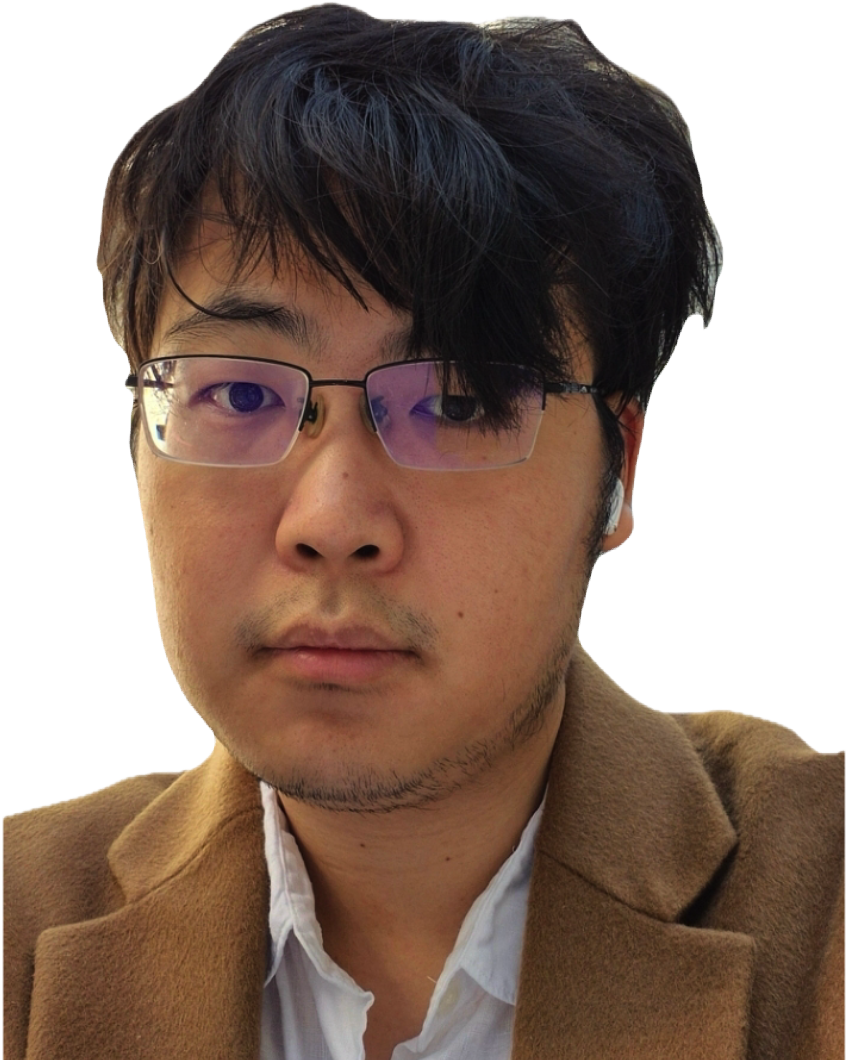}}]{Zhongchun Zhou}
 received the B.E. degree in microelectronic engineering from the School of Integrated Circuits, Tsinghua University, Beijing, China, in 2022. He is currently pursuing the Ph.D. degree in electronic and computer engineering with the Hong Kong University of Science and Technology, Hong Kong. His research interests include integrated circuits, field-programmable gate arrays (FPGAs), high performance computing (HPC) and computer architecture, with a focus on hardware acceleration and system optimization for large language models.
\end{IEEEbiography}

\begin{IEEEbiography}[{\includegraphics[width=1in,height=1.25in,clip,keepaspectratio]{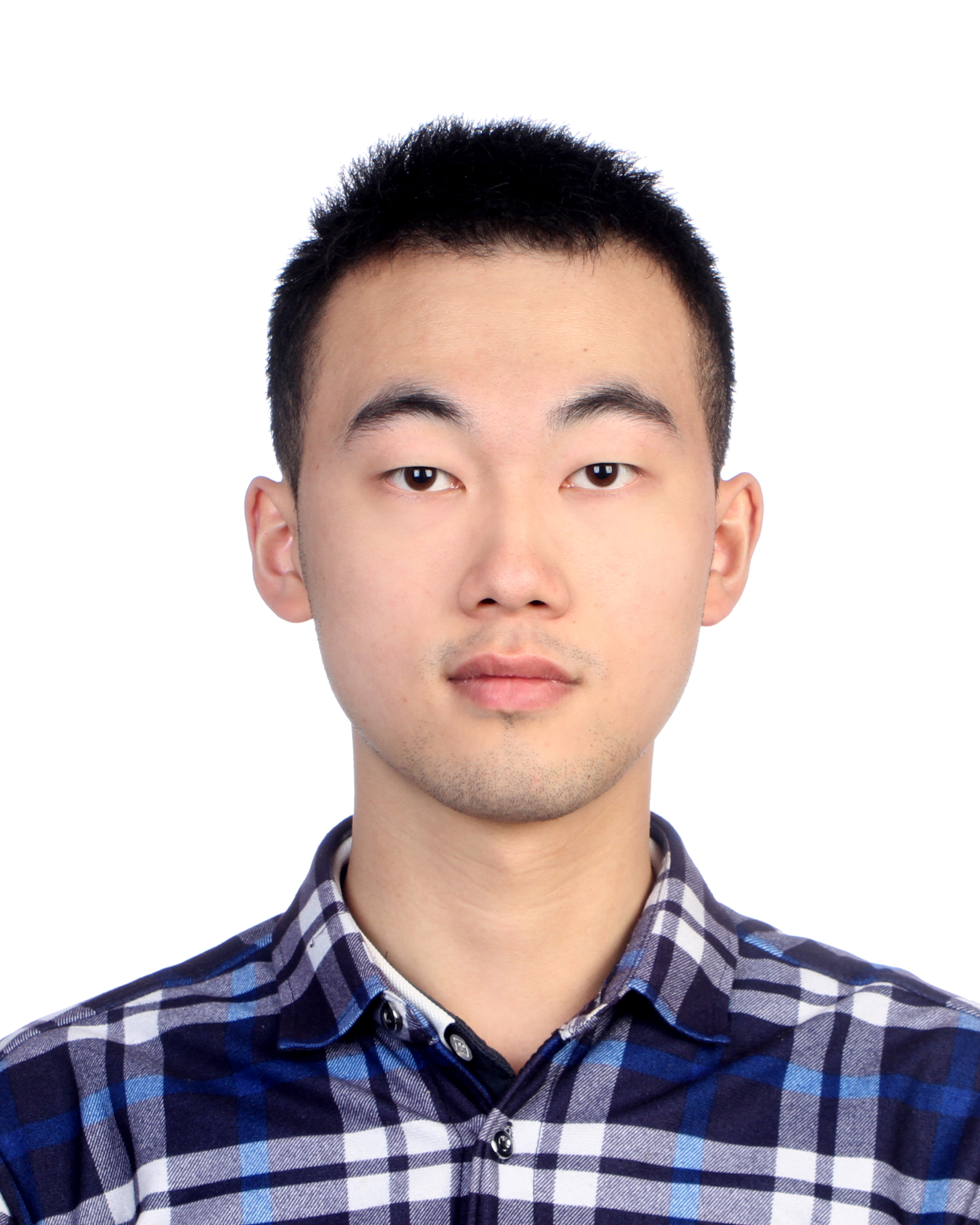}}]{Chengtao Lai}
 received the B.E. degree in Engineering Mechanics from the School of Aerospace Engineering, Tsinghua University, Beijing, China, in 2021. He is currently pursuing the Ph.D. degree in Electronic and Computer Engineering with the Hong Kong University of Science and Technology, Hong Kong. His research interests include high performance computing and computer architecture, especially the memory subsystem of AI accelerators.
\end{IEEEbiography}

\begin{IEEEbiography}[{\includegraphics[width=1in,height=1.25in,clip,keepaspectratio]{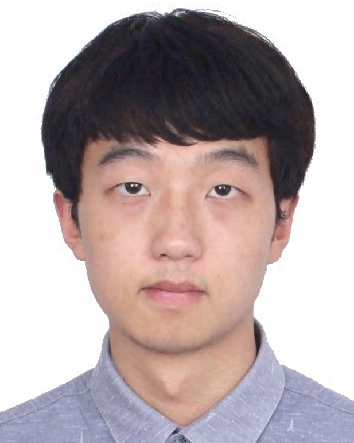}}]{Yuhang Gu}
received the B.E. degree in Electronic Science and Technology from the School of Electronic Science and Engineering, Southeast University, Nanjing, China, in 2026. He is currently pursuing the Ph.D. degree in electronic and computer engineering with the Hong Kong University of Science and Technology, Hong Kong. His research interests include computer systems and computer architecture.
\end{IEEEbiography}
\begin{IEEEbiography}[{\includegraphics[width=1in,height=1.25in,clip,keepaspectratio]{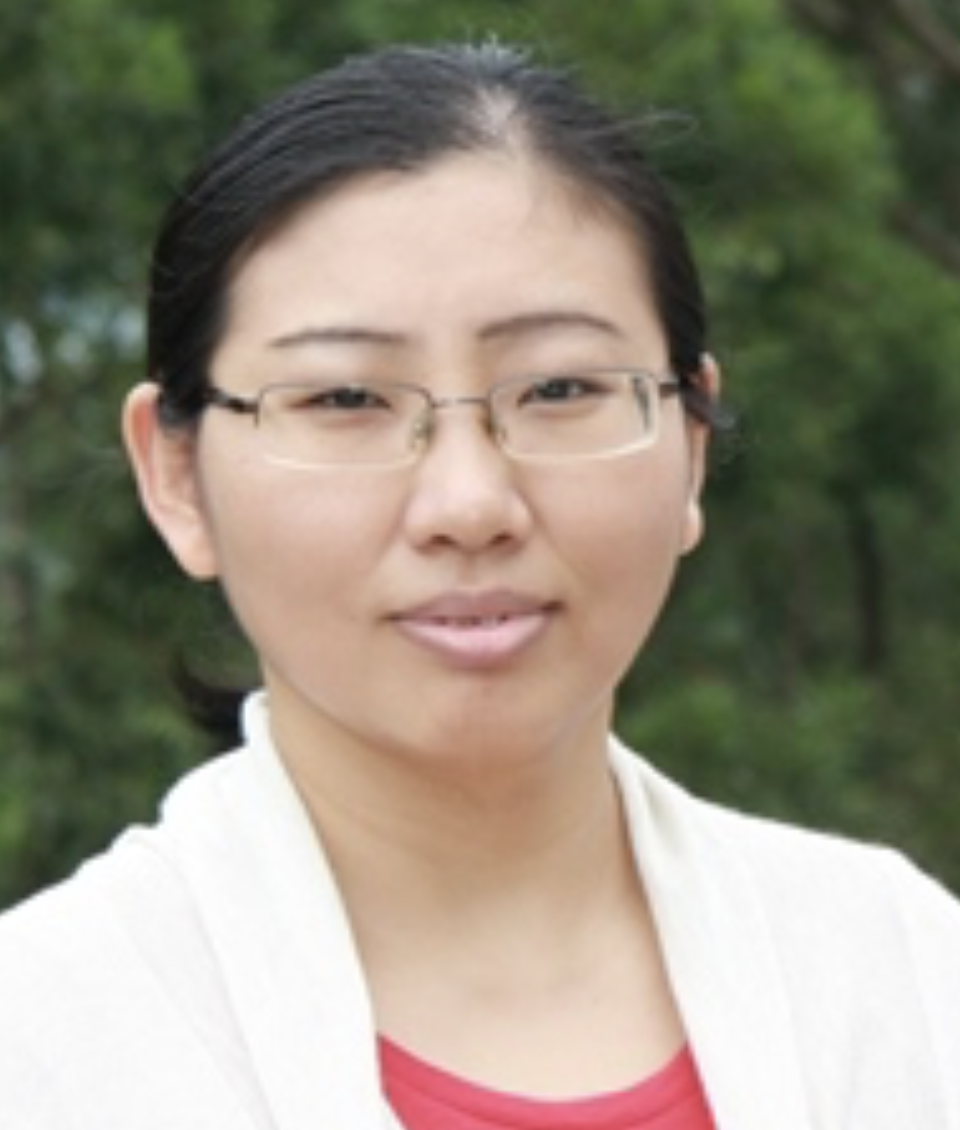}}]{Wei Zhang}
(Fellow, IEEE) received the Ph.D. degree from Princeton University, Princeton, NJ, USA, in 2009. She was an Assistant Professor with the School of Computer Engineering, Nanyang Technological University, Singapore, from 2010 to 2013. She joined the Hong Kong University of Science and Technology, Hong Kong, in 2013, where she is currently a Professor and she established the Reconfigurable Computing System Laboratory. She has authored or coauthored over 80 book chapters and papers in peer-reviewed journals and international conferences. Her current research interests include reconfigurable systems, FPGA-based design, low-power high-performance multicore systems, electronic design automation, embedded systems, and emerging technologies. Dr. Zhang serves as an Associate Editor for ACM Transactions on Embedded Computing Systems, IEEE Transactions on Very Large Scale Integration (VLSI) Systems, and ACM Journal of Emerging Technologies in Computing Systems. She also serves on many organization committees and technical program committees.
\end{IEEEbiography}




\end{document}